\documentclass[review]{elsarticle}
\pdfoutput=1

\PassOptionsToPackage{hyphens}{url}\usepackage{hyperref}
\usepackage[T1]{fontenc}%

\journal{Computer Physics Communications}

\usepackage{booktabs}
\usepackage{graphicx}

\usepackage{amsmath}
\DeclareMathOperator{\sign}{sign}

\usepackage{fancyvrb}
\usepackage{color}
\usepackage{listings}
\definecolor{lstcolstr}{rgb}{0,0.8,0}
\definecolor{lstcolkw}{rgb}{0,0,0.8}
\definecolor{lstcolcmt}{rgb}{0.8,0,0}
\lstdefinelanguage[mcpl]{C}[ANSI]{C}{
  morekeywords=[1]{
    mcpl_file_t, mcpl_outfile_t, mcpl_particle_t,
    int32_t, uint32_t, int64_t, uint64_t, size_t
}}
\lstdefinelanguage[mcpl]{C++}[]{C++}{
  morekeywords=[1]{
    mcpl_file_t, mcpl_outfile_t, mcpl_particle_t,
    int32_t, uint32_t, int64_t, uint64_t, size_t,
    G4String, G4int, G4double, G4ThreeVector
}}
\usepackage{lscape}
\usepackage{afterpage}
\usepackage{moresize}

\lstdefinelanguage[mccode]{C}[ANSI]{C}{
  morekeywords=[1]{
    DECLARE, DEFINE, END, FINALLY, INITIALIZE, MCDISPLAY, SAVE, SHARE,
    TRACE, DEFINITION, PARAMETERS, POLARISATION, SETTING,
    OUTPUT, INSTRUMENT, include,
    ABSOLUTE,AT,COMPONENT,EXTEND,GROUP,PREVIOUS,NEXT,MYSELF,RELATIVE,ROTATED,WHEN,JUMP,ITERATE,SPLIT,COPY
}}
\usepackage[range-units=single,range-phrase=--]{siunitx}

\newcommand{\fctname}[1]{{\lstinline{#1}}}%
\lstnewenvironment{docfct}
    {\lstset{language={[mcpl]C},
        basicstyle={\linespread{0.9}\ttfamily\scriptsize},
        belowskip=0em,
        aboveskip=\bigskipamount
      }}{}
\usepackage{changepage}
\newcommand{\fctdesc}[1]{%
  \begin{adjustwidth}{0.05\textwidth}{}
    \scriptsize
    #1
  \end{adjustwidth}
}

\makeatletter
\renewcommand\appendix{\par
  \setcounter{section}{0}%
  \setcounter{subsection}{0}%
  \setcounter{equation}{0}%
  \setcounter{table}{0}%
  \setcounter{figure}{0}%
  \setcounter{lstlisting}{0}%
  \gdef\theequation{\@Alph\c@section.\arabic{equation}}%
  \gdef\thefigure{\@Alph\c@section.\arabic{figure}}%
  \gdef\thetable{\@Alph\c@section.\arabic{table}}%
  \gdef\thesection{\appendixname\@Alph\c@section}%
  \gdef\thelstlisting{\@Alph\c@section.\arabic{lstlisting}}%
  \@addtoreset{equation}{section}%
  \@addtoreset{table}{section}%
  \@addtoreset{figure}{section}%
  \@addtoreset{lstlisting}{section}%
  \gdef\theHtable{\@arabic\c@section.\@arabic\c@table}%
  \gdef\theHfigure{\@arabic\c@section.\@arabic\c@figure}%
  \gdef\theHsection{Appendix.\@arabic\c@section}%
}
\makeatother

\begin{document}

\begin{frontmatter}

\title{Monte Carlo Particle Lists: MCPL}
\author[addressess]{T~Kittelmann\corref{mycorrespondingauthor}}
\cortext[mycorrespondingauthor]{Corresponding author. {\em Email address:} \texttt{thomas.kittelmann@esss.se}}

\author[addressdtunutech]{E~Klinkby}
\author[addressdtuphys]{E~B~Knudsen}
\author[addressdtuphys,addressess]{P~Willendrup}
\author[addressess,addressdtunutech]{X~X~Cai}
\author[addressess]{K~Kanaki}
\address[addressess]{European Spallation Source ERIC, Sweden}
\address[addressdtunutech]{DTU Nutech, Technical University of Denmark, Denmark}
\address[addressdtuphys]{DTU Physics, Technical University of Denmark, Denmark}

\begin{abstract}
  A binary format with lists of particle state information, for interchanging
  particles between various Monte Carlo simulation applications, is
  presented. Portable \texttt{C} code for file manipulation is made available to
  the scientific community, along with converters and plugins for several
  popular simulation packages.
\end{abstract}

\end{frontmatter}

{\bf PROGRAM SUMMARY}

\begin{small}
\noindent
{\em Manuscript Title:} Monte Carlo Particle Lists: MCPL                                 \\
{\em Authors:} T~Kittelmann, E~Klinkby, E~B~Knudsen, P~Willendrup, X~X~Cai and K~Kanaki \\
{\em Program Title:} MCPL                                          \\
{\em Journal Reference:}                                      \\
{\em Catalogue identifier:}                                   \\
{\em Licensing provisions:} CC0 for core MCPL, see LICENSE file for details.\\
{\em Programming language:} \texttt{C} and \texttt{C++}                                  \\
{\em Operating system:} Linux, OSX, Windows                        \\
{\em Keywords:}  \\
\texttt{MCPL}, Monte Carlo, particles, data storage, simulation, interchange
format, \texttt{C++}, \texttt{C}, \texttt{Geant4}, \texttt{MCNP}, \texttt{McStas}, \texttt{McXtrace}\\
{\em Classification:} 4, 9, 11, 17 \\
{\em External routines/libraries:} \texttt{Geant4}, \texttt{MCNP}, \texttt{McStas}, \texttt{McXtrace}           \\
{\em Nature of problem:}\\
Persistification of particle states in Monte Carlo simulations, for interchange
between simulation packages or for reuse within a single package.
   \\
{\em Solution method:}\\
  Binary interchange format with associated code written in portable \texttt{C} along with
  tools and interfaces for relevant simulation packages.
   \\
\end{small}

\section{Introduction}

The usage of Monte Carlo simulations to study the transport and interaction of
particles and radiation is a powerful and popular technique, finding use
throughout a wide range of fields -- including but not limited to both high
energy and nuclear physics, as well as space and medical
sciences~\cite{mc_app_radphys_2000}. Naturally, a plethora of different
frameworks and applications exist for carrying out these simulations
(cf.\ section~\ref{sec:plugins} for examples), with implementations in different
languages and domains ranging from general purpose to highly specialised field-
and application-specific.

A common principle used in the implementation of these applications is the
representation of particles by a set of state parameters -- usually including at
least particle type, time coordinate, position and velocity or momentum vectors
-- and a suitable representation of the geometry of the problem (either via
descriptions of actual surfaces and volumes in a virtual three-dimensional
space, or through suitable parameterisations). In the simplest scenario where no
variance-reduction techniques are employed, simulations are typically carried
out by proceeding iteratively in steps from an initial set of particles states,
with the state information being updated along the way as a result of the
pseudo-random or deterministic modelling of processes affecting the
particle. The modelling can represent particle self-interactions, interactions
with the material of the simulated geometry, or simply its forward transport
through the geometry, using either straight-forward ray-tracing techniques or
more complicated trajectory calculations as appropriate.  In addition to a
simple update of state parameters, the modelling can result in termination of
the simulation for the given particle or in the creation of new secondary
particle states, which will in turn undergo simulation themselves.

Occasionally, use-cases arise in which it would be beneficial to be able to
capture a certain subset of particle states present in a given simulation, in
order to continue their simulation at a later point in either the same or a
different framework. Such capabilities have typically been implemented using
custom application-specific means of data exchange, often involving the tedious
writing of custom input and output hooks for the specific frameworks and
use-cases in question. Here is instead presented a standard format for exchange
of particle state data, \emph{Monte Carlo Particle Lists} (\texttt{MCPL}), which
is intended to replace the plethora of custom converters with a more convenient
scenario in which experts of each framework implement converters to the common
format, as a one-time effort. The idea being that users of the various
frameworks then gain the ability to simply activate those pre-existing and
validated converters in order to carry out their work.

The present work originated in the needs for simulations at neutron scattering
facilities, where a multitude of simulation frameworks are typically used to
describe the various components from neutron production to detection, but
historically other conceptually similar formats have been and are used in high
energy physics to communicate particle states between event generators and
detector simulations~\cite{hepevt1989,hepmc2000,leshoucheseventfiles}. However,
these formats were developed for somewhat different purposes than the one
presented here, keeping simulation histories, focusing on the description of
intermediate unphysical or bound particles, existing primarily in-memory rather
than on-disk, or implemented in languages not readily accessible to applications
based on different technologies. For instance, \cite{hepevt1989} is defined as
an in-memory \texttt{FORTRAN} common block, \cite{hepmc2000} provides a
\texttt{C++} infrastructure for in-memory data with customisable
persistification, and \cite{leshoucheseventfiles} defines a text-based format
focused on descriptions of intermediate particles and lacking particle
positions. These existing solutions were thus deemed unfit for the
goals of the work presented here: a compact yet flexible on-disk binary format
for particle state information, portable, well-defined and able to accommodate a wide range of use-cases
with close to optimal storage requirements. The accompanying code with which to
access and manipulate the files should be small, efficient and easily integrated into
existing codes and build systems. Consequently, it was chosen to implement the
format through a set of \texttt{C} functions declared in a single header file,
\texttt{mcpl.h}, and implemented in a single file, \texttt{mcpl.c}. These two
files will here be referred to as the \emph{core} \texttt{MCPL} code, and are
made freely available under the CC0 1.0 Universal Creative Commons
license. Along with associated code examples, documentation, configuration files
(cf.\ section~\ref{sec:buildanddeploy}) and application-specific interface code
which is not embedded in the relevant upstream projects
(cf.\ sections~\ref{sec:plugins_geant4} and \ref{sec:plugins_mcnp}), these files
constitute the \texttt{MCPL} distribution. The present text concerns the second
public release of \texttt{MCPL}, version 1.1.0.  Future updates to the
distribution will be made available at the project website~\cite{mcplwww}.

\section{The \texttt{MCPL} format}

\texttt{MCPL} is a binary file format in which a header section, with
configuration and meta-data, is followed by a data section, where the state
information of the contained particles is kept. Data compression is available
but optional (cf.\ section~\ref{sec:compression}). The uncompressed storage size
of a particle entry in the data section is determined by overall settings in the
header section, and depends on what exact information is stored for the
particles in a given file, as will be discussed shortly. Within a given file,
all particle entries will always be of equal length, allowing for trivial
calculation of the absolute data location for a particle at a given index in the
file -- and thus for efficient seeking and skipping between particles if
desired. It is expected and recommended that \texttt{MCPL} files will be
manipulated, directly or indirectly, by calls to the functions in
\texttt{mcpl.h} (cf.\ section~\ref{sec:progmcplaccess}), but for reference a
complete specification of the binary layout of data in the files is provided in
\ref{appendix:mcpldetailedlayout}.

\subsection{Information available}\label{sec:format_infoavail}

\begin{table}[tbp]
\centering
\resizebox{\textwidth}{!}{%
\begin{tabular}{@{}llp{0.8\textwidth}@{}}
\toprule
\multicolumn{3}{c}{\textbf{File header information}}                                                                                                                                                               \\ \midrule
\textit{Field}                            &  & \textit{Description}                                                                                                                                           \\ \cmidrule(r){1-1} \cmidrule(l){3-3}
File type magic number 0x4d43504c ("MCPL")  &  & All \texttt{MCPL} files start with this 4-byte word.                                                                                                                    \\
Version                       &  & File format version.                                                                                                                                            \\
Endianness                                &  & Whether numbers in file are in little- or big-endian format.                                                                                                    \\
Number of particles in file               &  & 64 bit integer.                                                                                                                                                 \\
Flag : Particles have polarisation info   &  & If false, all loaded particles will have polarisation vectors (0,0,0).                                                                                          \\
Flag : Particles have user-flags field    &  & If false, all loaded particles will have user-flags 0x00000000.                                                                                                  \\
Flag : Particle info use double-precision &  & If true, floating point storage use double-precision.                                                                                                          \\
Global PDG code                           &  & If this 32 bit integer is non-zero, all loaded particles will have this PDG code.                                                                                \\
Global weight                             &  & If this double-precision floating point number is non-zero, all loaded particles will have this weight.                                                                                \\
Source name                               &  & String indicating the application which created the \texttt{MCPL} file.                                                                                                  \\
Comments                                  &  & A variable number of comments (strings) added at file creation.                                                                                                \\
Binary blobs                              &  & A variable number of binary data blobs, indexed by keys (strings). This allows arbitrary custom data to be embedded.\\ \bottomrule
\end{tabular}%
}
\caption{Information available in the header section of \texttt{MCPL} files.}
\label{tab:mcplhdr}
\end{table}

The information available in the file header is indicated in
Table~\ref{tab:mcplhdr}: a unique 4-byte magic number identifying the format
always starts all files, and is followed by the format version, the endianness
(\emph{little} or \emph{big}) in which numbers in the file are stored, and the
number of particles in the file. The versioning provides a clear path for future
updates to the format, without losing the ability to read files created with
previous versions of the \texttt{MCPL} code, and the endianness information
prevents interpretation errors on different machines (although at present, most
consumer platforms are little-endian).\footnote{In the current implementation,
  reading a little-endian \texttt{MCPL} file on a big-endian machine or vice
  versa triggers an error message. It is envisioned that a future version of the
  \texttt{MCPL} code could instead transparently correct the endianness at load
  time.} Next come five options indicating what data is stored per-particle,
which will be discussed in the next paragraph. Finally, the header contains
several options for embedding custom free-form information: first of all, the
source name, in the form of a single string containing the name and perhaps
version of the application which created the file. Secondly, any number of
strings can be added as human readable comments, and, thirdly, any number of
binary data blobs can be added, each identified by a string key. The
\texttt{MCPL} format itself provides no restrictions on what data, if any, can
be stored in these binary blobs, but useful content could for instance be a copy
of configuration data used by the source application when the given file was
produced, kept for later reference. Also note that, for reasons of security, no
code in the \texttt{MCPL} distribution ever attempts to interpret contents
stored in such binary data blobs.

\begin{table}[tbp]
\centering
\resizebox{\textwidth}{!}{%
\begin{tabular}{@{}llllc@{}}
\toprule
\multicolumn{5}{c}{\textbf{Particle state information}} \\ \midrule
\textit{Field}  &  & \textit{Description} &  & \textit{Bytes of storage used per entry (FP = 4 or 8 bytes)} \\ \cmidrule(r){1-1} \cmidrule(lr){3-3}\cmidrule(l){5-5}
PDG code       & & 32 bit integer indicating particle type. & & 0 or 4 \\
Position       & & Vector, values in centimetres.           & & 3FP \\
Direction      & & Unit vector along the particle momentum. & & 2FP \\
Kinetic energy & & Value in MeV.                            & & 1FP \\
Time           & & Value in milliseconds.                   & & 1FP \\
Weight         & & Weight or intensity.                     & & 0 or 1FP \\
Polarisation   & & Vector.                                  & & 0 or 3FP \\
User-flags     & & 32 bit integer with custom info. & & 0 or 4 \\ \bottomrule
\end{tabular}%
}
\caption{Particle state information available and uncompressed storage requirements for each entry in the data section of \texttt{MCPL} files.}
\label{tab:mcplpart}
\end{table}

Table~\ref{tab:mcplpart} shows the state information available per-particle in
\texttt{MCPL} files, along with the storage requirements of each field. Particle
position, direction, kinetic energy and time are always stored.\footnote{Note
  that a valid alternative to storing the directional unit vector along with the
  kinetic energy would have been the momentum vector. However, the choice here
  is consistent with the variables used in interfaces of both \texttt{MCNP} and
  \texttt{Geant4}, and means that the \texttt{mcpl2ssw} converter discussed in
  section~\ref{sec:plugins_mcnp} can be implemented without access to an
  unwieldy database of particle and isotope masses.} Polarisation vectors and
so-called \emph{user-flags} in the form of unsigned 32 bit integers are only
stored when relevant flags in the header are enabled and weights are only stored
explicitly in each entry when no global common value was set in the
header. Likewise, the particle type information in the form of so-called PDG
codes is only stored when a global PDG code was not specified in the header. The
PDG codes must follow the scheme developed by the Particle Data Group
in~\cite[ch.~42]{pdg2014}, which is inarguably the most comprehensive and widely
adopted standard for particle type encoding in simulations. Finally, again
depending on a flag in the header, particle information uses either single- (4
bytes) or double-precision (8 bytes) storage for floating point numbers. All in
all, summing up the numbers in the last column of Table~\ref{tab:mcplpart},
particles are seen to consume between 28 and 96 bytes of uncompressed storage
space per entry. The \texttt{MCPL} format is thus designed to be flexible enough
to handle use-cases requiring a high level of detail in the particle state
information, without imposing excessive storage requirements on less demanding
scenarios.

Note that while the units for position, energy and time indicated in
Table~\ref{tab:mcplpart} of course must be respected, the choices themselves are
somewhat arbitrary and should in no way be taken to indicate the suitability of
the \texttt{MCPL} format for a given simulation task. In particular, note that
within the dynamic range of a given floating point representation, the relative
numerical precision is essentially independent of the magnitude of the numbers
involved and is determined by the number of bits allocated for the
\emph{significand}~\cite{IEEE754}. Thus, it is important to realise that
usage of the \texttt{MCPL} format to deal with a simulation task whose natural
units are many orders of magnitude different than the ones in
Table~\ref{tab:mcplpart} does \emph{not} imply any detrimental impact on
numerical precision.

Packing of the three-dimensional unit directional vector into just two floating
point numbers of storage is carried out via a new packing algorithm, tentatively
named \emph{Adaptive Projection Packing}, discussed in detail in
\ref{appendix:unitvectorpacking}. Unlike other popular packing strategies
considered, the chosen algorithm provides what is for all practical purposes
flawless performance, with a precision comparable to the one existing absent any
packing (i.e.\ direct storage of all coordinates into three floating point
numbers). It does so without suffering from domain validity issues, and the
implemented code is not significantly slower to execute than the alternatives.

\subsection{Accessing or creating \texttt{MCPL} files programmatically}\label{sec:progmcplaccess}

While a complete documentation of the programming API provided by the
implementation of \texttt{MCPL} in \texttt{mcpl.h} and \texttt{mcpl.c} can be
found in \ref{appendix:reference_c_api}, the present discussion will restrict
itself to a more digestible overview.

The main feature provided by the API is naturally the ability to create new
\texttt{MCPL} files and access the contents of existing ones, using a set of
dedicated functions. No matter which settings were chosen when a given
\texttt{MCPL} file was created, the interface for accessing the header and
particle state information within it is the same, as can be seen in
Listing~\ref{lst:readexample}: after obtaining a file handle via
\texttt{mcpl\_open\_file}, a pointer to an \texttt{mcpl\_particle\_t}
\texttt{struct}, whose fields contain the state information available for a
given particle, is returned by calling \texttt{mcpl\_read}. This also advances
the position in the file, and returns a null-pointer when there are no more
particles in the file, ending the loop. If a file was created with either
polarisation vectors or user-flags disabled, the corresponding fields on the
particle will contain zeroes (thus representing polarisation information with
null-vectors and user-flags with an integer with no bits enabled). All floating
point fields on \texttt{mcpl\_particle\_t} are represented with a
double-precision type, but the actual precision of the numbers will obviously be
limited to that stored in the input file. In addition to the interface
illustrated by Listing~\ref{lst:readexample}, functions can be found in
\texttt{mcpl.h} for accessing any information available in the file header (see
Table~\ref{tab:mcplhdr}), or for seeking and skipping to particles at specific
positions in the file, rather than simply iterating through the full file.

\begin{lstlisting}[float,language={[mcpl]C},
  label={lst:readexample},
  caption={Simple example for looping over all particles in an existing \texttt{MCPL} file.}
]
#include "mcpl.h"

void example()
{
  mcpl_file_t f = mcpl_open_file("myfile.mcpl");

  const mcpl_particle_t* p;

  while ( ( p = mcpl_read(f) ) ) {

    /* Particle properties can here be accessed
       through the pointer "p":

       p->pdgcode
       p->position[k] (k=0,1,2)
       p->direction[k] (k=0,1,2)
       p->polarisation[k] (k=0,1,2)
       p->ekin
       p->time
       p->weight
       p->userflags
    */

  }

  mcpl_close_file(f);
}
\end{lstlisting}

Code creating \texttt{MCPL} files is typically slightly more involved, as the
creation process also involves deciding on the values of the various header
flags and filling of free-form information like source name and comments. An
example producing a file with 1000 particles is shown in
Listing~\ref{lst:writeexample}. The first part of the procedure is to obtain a
file handle through a call to \texttt{mcpl\_create\_outfile}, configure the header
and overall flags, and prepare a zero-initialised instance of
\texttt{mcpl\_particle\_t}. Next comes the loop filling the particles into the
file, which happens by updating the state information on the
\texttt{mcpl\_particle\_t} instance as needed, and passing it to
\texttt{mcpl\_add\_particle} each time. At the end, a call to
\texttt{mcpl\_close\_outfile} finishes up by flushing all internal buffers to
disk and updating the field containing the number of particles at the beginning
of the file.

\begin{lstlisting}[float,language={[mcpl]C},
  label={lst:writeexample},
  caption={Simple example for creating an \texttt{MCPL} file with 1000 particles.}
]
#include "mcpl.h"

void example()
{
  mcpl_outfile_t f = mcpl_create_outfile("myfile.mcpl");
  mcpl_hdr_set_srcname(f,"MyAppName-1.0");

  /* Tune file options or add custom comments or
     binary data into the header:

     mcpl_enable_userflags(f);
     mcpl_enable_polarisation(f);
     mcpl_enable_doubleprec(f);
     mcpl_enable_universal_pdgcode(f,myglobalpdgcode);
     mcpl_enable_universal_weight(f,myglobalweight);
     mcpl_hdr_add_comment(f,"Some comment.");
     mcpl_hdr_add_data(f,"mydatakey",
                       my_datalength, my_databuf)
  */

  mcpl_particle_t* p = mcpl_get_empty_particle(f);

  int i;
  for ( i = 0; i < 1000; ++i ) {

    /* The following particle properties must
       always be set here:

       p->position[k] (k=0,1,2)
       p->direction[k] (k=0,1,2)
       p->ekin
       p->time

       These should also be set when required by
       file options:

       p->pdgcode
       p->weight
       p->userflags
       p->polarisation[k] (k=0,1,2)
    */

    mcpl_add_particle(f,p);
  }

  mcpl_close_outfile(f);
}
\end{lstlisting}

Should the program abort before the call to
\texttt{mcpl\_close\_outfile}, particles already written into the output file
are normally recoverable: upon opening such an incomplete file, the
\texttt{MCPL} code detects that the actual size of the file is inconsistent with
the value of the field in the header containing the number of particles. Thus,
it emits a warning message and calculates a more appropriate value for the
field, ignoring any partially written particle state entry at the end of the
file. This ability to transparently correct incomplete files upon load also
means that it is possible to inspect (with the \texttt{mcpltool} command
discussed in section~\ref{sec:mcplfileaccesscmdline}) or analyse files that are
still being created. To avoid seeing a warning each time a file left over from
an aborted job is opened, \texttt{mcpl.h} also provides the function
\texttt{mcpl\_repair} which can be used to permanently correct the header of the
file.

Likewise, \texttt{mcpl.h} also provides the function \texttt{mcpl\_merge\_files} which
can be used to merge a list of compatible \texttt{MCPL} files into a new one, which might
typically be useful when gathering up the output of simulations carried out via
parallel processing techniques. Compatibility here means that the files must
have essentially identical header sections, except for the field holding the
number of particles. Finally, the function \texttt{mcpl\_transfer\_metadata} can
be used to easily implement custom extraction of particle subsets from existing
\texttt{MCPL} files into new (smaller) ones. An example of this is illustrated
in Listing~\ref{lst:editexample}.

\begin{lstlisting}[float,language={[mcpl]C},
  label={lst:editexample},
  caption={Example extracting low-energy neutrons (PDG code 2112) from an \texttt{MCPL} file.}
]
#include "mcpl.h"

void example() {

  /* open files, transfer meta-data, add comment */

  mcpl_file_t fi = mcpl_open_file("myfile.mcpl");
  mcpl_outfile_t fo = mcpl_create_outfile("new.mcpl");
  mcpl_transfer_metadata(fi, fo);
  mcpl_hdr_add_comment(fo,"Extracted neutrons with ekin<0.1MeV");

  /* transfer selected particles */

  const mcpl_particle_t* particle;
  while ( ( particle = mcpl_read(fi) ) ) {
    if ( particle->pdgcode == 2112 && particle->ekin < 0.1 )
      mcpl_add_particle(fo,particle);
  }

  /* finish up */

  mcpl_close_outfile(fo);
  mcpl_close_file(fi);
}

\end{lstlisting}

\subsection{Accessing \texttt{MCPL} files from the command line}\label{sec:mcplfileaccesscmdline}

Compared with simpler text-based formats (e.g.\ ASCII files with data formatted
in columns), one potential disadvantage of a binary data format like
\texttt{MCPL} is the lack of an easy way for users to quickly inspect a file and
investigate its contents. To alleviate this, \texttt{mcpl.h} provides a function
which, in a straight-forward manner, can be used to build a generic
\texttt{mcpltool} command-line executable:
\texttt{int~mcpl\_tool(int~argc,char**~argv)}, for which full usage instructions
can be found in \ref{appendix:reference_mcpltool_usage} or by invoking it with
the \texttt{-{}-help} flag. Simply running this command on an
\texttt{MCPL} file without specifying other arguments, results in a short
summary of the file content being printed to standard output, which includes a
listing of the first 10 contained particles. An example of such a summary is
provided in Listing~\ref{lst:mcpltool_example_output}: it is clear from the
displayed meta-data that the particles in the given file represent a
transmission spectrum resulting from illumination of a block of lead by a
\SI{10}{GeV} proton beam in a \texttt{Geant4}~\cite{geant4a,geant4b}
simulation. The displayed header information and data columns should be mostly
self-explanatory, noting that $(\texttt{x},\texttt{y},\texttt{z})$ indicates the
particle position, $(\texttt{ux},\texttt{uy},\texttt{uz})$ its normalised
direction, and that the \texttt{pdgcode} column indeed shows particle types
typical in a hadronic shower: $\pi^+$ (211), $\gamma$ (22), protons (2212),
$\pi^-$ ($-211$) and neutrons (2112). If the file had user-flags or polarisation
vectors enabled, appropriate columns for those would be shown as well. Finally,
note that the \texttt{36~bytes/particle} refers to uncompressed storage, and
that in this particular case the file actually has a compression ratio of
approximately 70\%, meaning that about 25 bytes of on-disk storage is used per
particle (cf.\ section~\ref{sec:compression}).

\afterpage{\begin{landscape}
\begin{lstlisting}[float,language={},basicstyle={\linespread{0.9}\ttfamily\ssmall},
  label={lst:mcpltool_example_output},
  caption={Example output of running \texttt{mcpltool} with no arguments on a
    specific \texttt{MCPL} file.}
]
Opened MCPL file myoutput.mcpl.gz:

  Basic info
    Format             : MCPL-3
    No. of particles   : 1106933
    Header storage     : 140 bytes
    Data storage       : 39849588 bytes

  Custom meta data
    Source             : "G4MCPLWriter [G4MCPLWriter]"
    Number of comments : 1
          -> comment 0 : "Transmission spectrum from 10GeV proton beam on 20cm lead"
    Number of blobs    : 0

  Particle data format
    User flags         : no
    Polarisation info  : no
    Fixed part. type   : no
    Fixed part. weight : no
    FP precision       : single
    Endianness         : little
    Storage            : 36 bytes/particle

index     pdgcode   ekin[MeV]       x[cm]       y[cm]       z[cm]          ux          uy          uz    time[ms]      weight
    0         211      487.02     -0.5898       1.835          20   -0.092407     0.20491     0.97441  7.3346e-07           1
    1          22      1.5326      1.0635      11.351          20    0.080441     0.66026     0.74672  1.0882e-06           1
    2          22      3.9526    -0.43907      8.6473          20    -0.56616     0.50558     0.65104  1.0286e-06           1
    3          22     0.82591      1.7444      9.7622          20    0.092099     0.79597     0.59829  1.0378e-06           1
    4          22      1.1958      2.1806      8.6416          20     0.21997     0.66435     0.71432  1.0124e-06           1
    5          22      1.2525      3.0949      7.7366          20     0.48903     0.30789     0.81612   1.013e-06           1
    6          22      2.6247       3.948       5.681          20     0.62503     0.64221     0.44374  9.1152e-07           1
    7        2212      824.28     -1.8797     -2.5124          20     -0.3077    -0.40496       0.861  7.6539e-07           1
    8        -211      3459.8    -0.79521     0.91481          20    -0.13441     0.14438     0.98035  7.0618e-07           1
    9        2112     0.30553      54.471      33.386          20      0.4862    0.011958     0.87377  0.00016442           1
\end{lstlisting}
\end{landscape}}

By providing suitable arguments (cf.~\ref{appendix:reference_mcpltool_usage}) to
\texttt{mcpltool}, it is possible to modify what information from the file is
displayed. This includes the possibility to change what particles from the file,
if any, should be listed, as well as the option to extract the contents of a
given binary data blob to standard output. The latter might be particularly
handy when entire configuration files have been embedded
(cf.\ sections~\ref{sec:plugins_mcnp} and
\ref{sec:plugins_mcstasmcxtrace}). Finally, the \texttt{mcpltool} command also
allows file merging and repairing, as discussed in
section~\ref{sec:progmcplaccess}, and provides functionality for selecting a
subset of particles from a given file and extracting them into a new smaller
file.

Advanced functionality such as graphics display and interactive GUI-based
investigation or manipulation of the contents of \texttt{MCPL} files is not
provided by the \texttt{mcpltool}, since those would imply additional unwanted
dependencies to the core \texttt{MCPL} code, which is required by design to be
light-weight and widely portable. However, it is the hope that the existence of
a standard format like \texttt{MCPL} will encourage development of such tools,
and indeed some already exist in the in-house framework~\cite{dgcodechep2013}
of the Detector Group at the European Spallation Source
(ESS)~\cite{esscdr,esstdr}. It is intended for a future distribution of
\texttt{MCPL} to include relevant parts of these tools as a separate and
optional component.

\subsection{Compression}\label{sec:compression}

The utilisation of data compression in a format like \texttt{MCPL} is
potentially an important feature, since on-disk storage size could be a concern
for some applications. Aiming to maximise flexibility, transparency and
portability, optional compression of \texttt{MCPL} files is simply provided by
allowing whole-file compression into the widespread \texttt{GZIP}
format~\cite{RFC1952_gzip} (changing the file extension from \texttt{.mcpl} to
\texttt{.mcpl.gz} in the process). This utilises the \texttt{DEFLATE}
compression algorithm~\cite{RFC1951_deflate} which offers a good performance
compromise with a reasonable compression ratio and an excellent speed of
compression and decompression.

Relying on a standard format such as \texttt{GZIP} means that, if needed, users
can avail themselves of existing tools (like the \texttt{gzip} and
\texttt{gunzip} commands available on most \texttt{UNIX} platforms) to change
the compression state of an existing \texttt{MCPL} file. However, when the code
in \texttt{mcpl.c} is linked with the ubiquitous
\texttt{ZLIB}~\cite{zlib_libandwww,RFC1950_zlib}
(cf.\ section~\ref{sec:buildanddeploy}), compressed \texttt{MCPL} files can be
read directly. For convenience, \texttt{mcpl.h} additionally provides a function
\texttt{mcpl\_closeandgzip\_outfile}, which can be used instead of
\texttt{mcpl\_close\_outfile} (cf.\ Listing~\ref{lst:writeexample}) to ensure
that newly created \texttt{MCPL} files are automatically compressed if possible
(either through a call to an external \texttt{gzip} command or through custom
\texttt{ZLIB}-dependent code, depending on availability).

\subsection{Build and deployment}\label{sec:buildanddeploy}

It is the hope that eventually \texttt{MCPL} capabilities will be included
upstream in many applications, and that users of those consequently won't have
to do anything extra to start using it. As will be discussed in
section~\ref{sec:plugins}, this is at present the case for users of recent
versions of \texttt{McStas}~\cite{mcstas1,mcstas2} and
\texttt{McXtrace}~\cite{mcxtrace1}, and is additionally the case for users of
the in-house \texttt{Geant4}-based framework of the ESS Detector
Group~\cite{dgcodechep2013}.

By design, it is expected that most developers wishing to add \texttt{MCPL}
support to their application will simply place copies of \texttt{mcpl.h} and
\texttt{mcpl.c} into their existing build system and include \texttt{mcpl.h}
from either \texttt{C} or \texttt{C++} code.\footnote{Compilation of
  \texttt{mcpl.c} can happen with any of the following standards: \texttt{C99},
  \texttt{C11}, \texttt{C++98}, \texttt{C++11}, \texttt{C++14}, or later. In
  addition to those, \texttt{mcpl.h} is also \texttt{C89} compatible. Note that
  on platforms where the standard \texttt{C} math function \texttt{sqrt} is
  provided in a separate library, that library must be available at link-time.}
In order to make the resulting binary code able to manipulate compressed files
directly (cf.\ section~\ref{sec:compression}), the code in \texttt{mcpl.c} must
usually be compiled against and linked with an installation of \texttt{ZLIB}
(see detailed instructions regarding build flags at the top of
\texttt{mcpl.c}). Alternatively, the \texttt{MCPL} distribution presented here
contains a ``fat'' auto-generated drop-in replacement for \texttt{mcpl.c} named
\texttt{mcpl\_fat.c}, in which the source code of \texttt{ZLIB} has been
included in its entirety.\footnote{Note that all \texttt{ZLIB} symbols have been
  prefixed, to guard against potential run-time clashes where a separate
  \texttt{ZLIB} is nonetheless loaded.} Using this somewhat larger file enables
\texttt{ZLIB}-dependent code in \texttt{MCPL} even in situations where
\texttt{ZLIB} might not be otherwise available.

In addition to the core \texttt{MCPL} code, the \texttt{MCPL} distribution also
contains a small file providing the \texttt{mcpltool} executable, \texttt{C++}
files implementing the \texttt{Geant4} classes discussed in
section~\ref{sec:plugins_geant4}, \texttt{C} files for the \texttt{mcpl2ssw} and
\texttt{ssw2mcpl} executables discussed in section~\ref{sec:plugins_mcnp}, and a
few examples show-casing how user code might look.

Building of all of these parts should be straight-forward using standard tools,
but a configuration file for \texttt{CMake}~\cite{cmakebook2015} which builds
and installs everything is nonetheless provided for reference and
convenience. Additionally, ``fat'' single-file versions of all command line
utilities (\texttt{mcpltool}, \texttt{mcpl2ssw} and \texttt{ssw2mcpl}) are also
provided, containing both \texttt{MCPL} and \texttt{ZLIB} code within as
appropriate. Thus, any of these single-file versions can be compiled directly
into the corresponding command line executable, without any other dependencies
than a \texttt{C} compiler. For more details about how to build and deploy,
refer to the \texttt{INSTALL} file shipped with the \texttt{MCPL} distribution.

\section{Application-specific converters and plugins}\label{sec:plugins}

While the examples in section~\ref{sec:progmcplaccess} show how it is possible
to manipulate \texttt{MCPL} files directly from \texttt{C} or \texttt{C++} code,
it is not envisioned that most users will have to write such code
themselves. Rather, in addition to using available tools (such as the
\texttt{mcpltool} described in section~\ref{sec:mcplfileaccesscmdline}) to
access the contents of files as needed, users would ideally simply use
pre-existing plugins and converters written by application-specific experts, to
load particles from \texttt{MCPL} files into their given Monte Carlo
applications, or extract particles from those into \texttt{MCPL} files. At the
time of this initial public release of \texttt{MCPL}, four such applications are
already \texttt{MCPL}-aware in this manner: \texttt{Geant4}, \texttt{MCNP},
\texttt{McStas} and \texttt{McXtrace}, and the details of the corresponding
converters and plugins are discussed in the following sub-sections, after a few
general pieces of advice for other implementers in the next paragraphs.

In order for \texttt{MCPL} files to be as widely exchangeable as possible, code
loading particles from \texttt{MCPL} files into a given Monte Carlo application
should preferably be as accepting as possible. In particular, this means that
warnings rather than errors should result if the input file contains PDG codes
corresponding to particle types that can not be handled by the application in
question. As an example, a detailed \texttt{MCNP} or \texttt{Geant4} simulation
of a moderated neutron source will typically produce files containing not only
neutrons, but also gammas and other particles. It should certainly be possible
to load such a file into a neutron-only simulation application like
\texttt{McStas}, resulting in simulation of the contained neutrons (preferably
with a warning or informative message drawing attention to some particles being
ignored).

Applications employing parallel processing techniques, must always pay
particular attention when implementing file-based I/O, and this is naturally
also the case when creating \texttt{MCPL}-aware plugins for them. However, the
available functionality for merging of \texttt{MCPL} files makes the scenario of
file creation particularly simple to implement: each sub-task can simply write
its own file, with the subsequent merging into a single file taking place during
post-processing. For reading of particles in existing \texttt{MCPL} files, it is
recommended that each sub-task performs a separate call to
\texttt{mcpl\_open\_file}, and use the skipping and seeking functionality to
load just a subset of the particles within, as required. In the case of a
multi-threading application, it is of course also possible to handle concurrent
input or output directly through a single file handle. In this case, however,
calls to \texttt{mcpl\_add\_particle} and \texttt{mcpl\_read} must be protected
against concurrent invocations with a suitable lock or mutex.

The following three sub-sections are dedicated to discussions of presently
available \texttt{MCPL} interfaces for specific Monte Carlo applications. The
discussions will in each case presuppose familiarity with the application in
question.

\subsection{\texttt{Geant4} interface}\label{sec:plugins_geant4}

In the most typical mode of working with the
\texttt{Geant4}~\cite{geant4a,geant4b} toolkit, users create custom \texttt{C++}
classes, sub-classing appropriate abstract interfaces, in order to set up
geometry, particle generation, custom data readout and physics modelling. At
run-time, those classes are then instantiated and registered with the
framework. Accordingly, the \texttt{MCPL}--\texttt{Geant4} integration takes the
form of two such sub-classes of \texttt{Geant4} interface classes, which can be
either directly instantiated or further sub-classed themselves as needed:
\texttt{G4MCPLGenerator} and \texttt{G4MCPLWriter}. They are believed to be
compatible with any recent version of \texttt{Geant4} and were explicitly tested
with versions 10.00.p03 and 10.02.p02.

First, the \texttt{G4MCPLGenerator}, the relevant parts of which are shown in
Listing~\ref{lst:g4mcplgenerator}, implements a \texttt{Geant4} generator by
sub-classing the \texttt{G4VUser\-Primary\-Generator\-Action} interface class. The
constructor of \texttt{G4MCPLGenerator} must be provided with the path to an
\texttt{MCPL} file, which will then be read one particle at a time whenever
\texttt{Geant4} calls the \texttt{GeneratePrimaries} method, in order to
generate \texttt{Geant4} events with a single primary particle
in each. If the file runs out of particles before the \texttt{Geant4} simulation
is ended for other reasons, the \texttt{G4MCPLGenerator} graciously requests the
\texttt{G4RunManager} to abort the simulation. Thus, a convenient way in which
to use the entire input file for simulation is to launch the simulation with a
very high number of events requested, as is done in the example in
Listing~\ref{lst:example_g4gen}.\footnote{Unfortunately, due to a limitation in
  the \texttt{G4RunManager} interface, this number will be limited by the
  highest number representable with a \texttt{G4int}, which on most modern
  platforms is 2147483647.}

\begin{lstlisting}[float,language={[mcpl]C++},
  label={lst:g4mcplgenerator},
  caption={The \texttt{G4MCPLGenerator} class.}
]
class G4MCPLGenerator : public G4VUserPrimaryGeneratorAction
{
  public:

    G4MCPLGenerator(const G4String& inputFile);
    virtual ~G4MCPLGenerator();
    virtual void GeneratePrimaries(G4Event*);

  protected:

    //Reimplement this to filter input particles (default
    //implementation accepts all particles):
    virtual bool UseParticle(const mcpl_particle_t*) const;

    //Reimplement this to change coordinates or weights of
    //input particles before using them (does nothing by
    //default):
    virtual void ModifyParticle(G4ThreeVector& position,
                                G4ThreeVector& direction,
                                G4ThreeVector& polarisation,
                                G4double& time,
                                G4double& weight) const;

  private:
    // ..
};
\end{lstlisting}

\begin{lstlisting}[float,language={[mcpl]C++},
  label={lst:example_g4gen},
  caption={Example showing how to load particles from an \texttt{MCPL} file into
    a \texttt{Geant4} simulation.}
]
#include "G4MCPLGenerator.hh"
#include "G4RunManager.hh"
#include <limits>

//Not shown here: Code defining MyGeometry and MyPhysicsList.

int main( int argc, char** argv ) {

  G4RunManager runManager;
  runManager.SetUserInitialization(new MyGeometry);
  runManager.SetUserInitialization(new MyPhysicsList);
  runManager.SetUserAction(new G4MCPLGenerator("myfile.mcpl"));
  runManager.Initialize();
  runManager.BeamOn(std::numeric_limits<G4int>::max());

  return 0;
}
\end{lstlisting}

In case the user wishes to use only certain particles from the input file for
simulation, the \texttt{G4MCPLGenerator} class must be sub-classed and the
\texttt{UseParticle} method reimplemented, returning \texttt{false} for
particles which should be skipped. Likewise, if it is desired to perform
coordinate transformations or reweighing before using the loaded particles, the
\texttt{ModifyParticle} method must be reimplemented.

The \texttt{G4MCPLWriter} class, the relevant parts of which are shown in
Listing~\ref{lst:g4mcplwriter}, is a \texttt{G4VSensitiveDetector} which in the
default configuration ``consumes'' all particles which, during a simulation,
enter any geometrical volume(s) to which it is attached by the user and stores
them into the specified \texttt{MCPL} file. At the same time it asks
\texttt{Geant4} to end further simulation of those particles (``killing''
them). This strategy of killing particles stored into the file was chosen as a
sensible default behaviour, as it prevents potential double-counting in the
scenarios where a particle (or its induced secondary particles) would otherwise
be able to enter a given volume multiple times. If it is desired to modify this
strategy, the user must sub-class \texttt{G4MCPLWriter} and reimplement the
\texttt{ProcessHits} method, using calls to \texttt{StorePreStep},
\texttt{StorePostStep} and \texttt{Kill}, as appropriate. For reference, code
responsible for the default implementation is shown in
Listing~\ref{lst:g4mcplwriter_processhits}. Likewise, to add \texttt{MCPL}
user-flags into the file, the \texttt{UserFlagsDescription} and
\texttt{UserFlags} methods must simply be reimplemented - the description
naturally ending up as a comment in the output file.

\begin{lstlisting}[float,language={[mcpl]C++},
  label={lst:g4mcplwriter},
  caption={The \texttt{G4MCPLWriter} class.}
]

class G4MCPLWriter : public G4VSensitiveDetector
{
  public:

    //Basic interface:

    G4MCPLWriter( const G4String& outputFile,
                  const G4String& name = "G4MCPLWriter" );
    virtual ~G4MCPLWriter();

    void AddComment( const G4String& );
    void AddData( const G4String& data_key,
                  size_t data_length,
                  const char* data );
    void EnableDoublePrecision();
    void EnablePolarisation();
    void EnableUniversalWeight(G4double);

    //Optional reimplement this to change default
    //"store-and-kill at entry" strategy:

    virtual G4bool ProcessHits( G4Step * step,
                                G4TouchableHistory* );

    //Optional reimplement these to add MCPL userflags:

    virtual G4String UserFlagsDescription() const { return ""; }
    virtual uint32_t UserFlags(const G4Step*) const { return 0; }


  protected:
    //Methods that can be used if reimplementing ProcessHits():
    void StorePreStep(const G4Step *);
    void StorePostStep(const G4Step *);
    void Kill(G4Step *);

  private:
    // ...
};
\end{lstlisting}

\begin{lstlisting}[float,language={[mcpl]C++},
  label={lst:g4mcplwriter_processhits},
  caption={The default \texttt{ProcessHits} implementation in the \texttt{G4MCPLWriter} class.}
]
G4bool G4MCPLWriter::ProcessHits(G4Step * step,G4TouchableHistory*)
{
  //Only consider particle steps originating at the boundary
  //of the monitored volume:
  if ( step->GetPreStepPoint()->GetStepStatus() != fGeomBoundary )
    return false;

  //Store the state at the beginning of the step, but avoid
  //particles taking their very first step (this would double-
  //count secondary particles created at the volume edge):
  if ( step->GetTrack()->GetCurrentStepNumber() > 1 )
    StorePreStep(step);

  //Tell Geant4 to stop further tracking of the particle:
  Kill(step);
  return true;
}
\end{lstlisting}

In Listing~\ref{lst:example_g4write} is shown how the \texttt{G4MCPLWriter} will
typically be configured and attached to logical volume(s) of the geometry.

\begin{lstlisting}[float,language={[mcpl]C++},
  label={lst:example_g4write},
  caption={Example showing how to produce an \texttt{MCPL} file from a \texttt{Geant4} simulation.}
]
//Provide output filename when creating G4MCPLWriter instance:
G4MCPLWriter * mcplwriter = new G4MCPLWriter("myoutput.mcpl");

//Optional calls which add meta-data or change flags:
mcplwriter->AddComment("Some useful description here");
mcplwriter->AddData( ... );
mcplwriter->EnableDoublePrecision();
mcplwriter->EnablePolarisation();
mcplwriter->EnableUniversalWeight(1.0);

//Register with G4SDManager and on one or more logical
//volumes to activate:
G4SDManager::GetSDMpointer()->AddNewDetector(mcplwriter);
alogvol->SetSensitiveDetector(mcplwriter);
\end{lstlisting}

\subsection{\texttt{MCNP} interface}\label{sec:plugins_mcnp}

Most users of \texttt{MCNP} are currently employing one of three distinct
flavours: \texttt{MCNPX}~\cite{mcnpx2006,mcnpx2011}, \texttt{MCNP5}~\cite{mcnp5}
or \texttt{MCNP6}~\cite{mcnp6}. In the most typical mode of working with any of
these software packages, users edit and launch \texttt{MCNP} through the use of
text-based configuration files (so-called \emph{input decks}), in order to set
up details of the simulation including geometry, particle generation, and data
extraction. The latter typically results in the creation of data files
containing simulation results, ready for subsequent analysis.

Although it would be conceivable to write in-process \texttt{FORTRAN}-compatible
\texttt{MCPL} hooks for \texttt{MCNP}, such an approach would require users to
undertake some form of compilation and linking procedure. This would likely
impose a change in working mode for the majority of \texttt{MCNP}
users, in addition to possibly requiring a special license for source-level
access to \texttt{MCNP}. Instead, the \texttt{MCNP}--\texttt{MCPL} interface
presented here exploits the existing \texttt{MCNP} capability to stop and
subsequently restart simulations at a user-defined set of surfaces through the
\texttt{Surface Source Write/Read} (\texttt{SSW}/\texttt{SSR}) functionality. As
the name suggests, the state parameters of simulated particles crossing a given
surface are stored on disk in dedicated files, with the intentional use as a
surface source in subsequent simulations with the same \texttt{MCNP}
setup. Presumably, these files (henceforth denoted ``\texttt{SSW} files'' in the
present text) are intended for this internal intermediate usage only, since their
format differs between different flavours of \texttt{MCNP}, and little effort
has been made to document the format in publicly available manuals. Despite
these obstacles, the \texttt{SSW} format is stable enough that several existing
\texttt{MCNP}-aware tools
(e.g.~\cite{Klinkby2013106,kbat_mctools_2015,pyne2014}) have chosen to provide
converters for this format, with various levels of functionality, and it was
thus deemed suitable also for the needs of the \texttt{MCPL} project.

Thus, the \texttt{MCPL} distribution presented here includes dependency-free
\texttt{C} code for two standalone executables, \texttt{mcpl2ssw} and
\texttt{ssw2mcpl}, which users can invoke from the command-line in order to
convert between \texttt{MCPL} and \texttt{SSW} files.\footnote{Prior work
  in~\cite{Klinkby2013106,kbat_mctools_2015} served as valuable input when
  developing code for interpreting data sections in \texttt{SSW} files.} The
usage of these two executables will be discussed here, while users are referred
to the relevant \texttt{MCNP} manuals for details of how to set up their input
decks to enable \texttt{SSW} input or output in their \texttt{MCNP} simulations:
\cite[Ch.~II.3.7]{mcnp5man}, \cite[Ch.~5.5.5]{mcnpxman} and
\cite[Ch.~3.3.4.7]{mcnp6man}. Note that through usage of \texttt{ssw2mcpl} and
\texttt{mcpl2ssw}, it is even possible to transfer particles between different
flavours and versions of \texttt{MCNP}, which is otherwise not possible with
\texttt{SSW} files.

First, the \texttt{ssw2mcpl} command, for which the full usage instructions are
shown in Listing~\ref{lst:ssw2mcplusage}, is in its most base invocation
straight-forward to use. Simply provide it with the name of an existing
\texttt{SSW} file to run on, and it will result in the creation of a new
(compressed) \texttt{MCPL} file, \texttt{output.mcpl.gz}, containing a copy of
all particles found in the \texttt{SSW} file. The \texttt{MCNP} flavour
responsible for creating the \texttt{SSW} file is automatically detected, the
resulting differences in the file format are taken into account behind the
scenes, and the detected \texttt{MCNP} version is documented as a comment in the
header of the resulting \texttt{MCPL} file.

\begin{lstlisting}[float,language={},
  basicstyle={\linespread{0.9}\ttfamily\scriptsize},
  label={lst:ssw2mcplusage},
  caption={Usage instructions for the \texttt{ssw2mcpl} command.}
]
Usage:

  ssw2mcpl [options] input.ssw [output.mcpl]

Converts the Monte Carlo particles in the input.ssw file (MCNP Surface
Source Write format) to MCPL format and stores in the designated output
file (defaults to "output.mcpl").

Options:

  -h, --help   : Show this usage information.
  -d, --double : Enable double-precision storage of floating point values.
  -s, --surf   : Store SSW surface IDs in the MCPL userflags.
  -n, --nogzip : Do not attempt to gzip output file.
  -c FILE      : Embed entire configuration FILE (the input deck)
                 used to produce input.ssw in the MCPL header.
\end{lstlisting}

The only relevant piece of information which is by default not transferred from
the \texttt{SSW} particle state into the \texttt{MCPL} file is the numerical ID
of the surface where the particle was registered in the \texttt{MCNP}
simulation. By supplying the \texttt{-s} option, \texttt{ssw2mcpl} will transfer
those to the \texttt{MCPL} user-flags field, and document this in the
\texttt{MCPL} header.  Additionally, while floating point numbers in the
\texttt{SSW} file are always stored in double-precision, the transfer to
\texttt{MCPL} will by default convert them to single-precision. This was chosen
as the default behaviour to keep usual storage requirements low, as
single-precision is arguably sufficient for most studies. By supplying the
\texttt{-d} option, \texttt{ssw2mcpl} will keep the numbers in double-precision
in the \texttt{MCPL} file as well. Depending on compression and the applied
flags, the on-disk size of the resulting \texttt{MCPL} file will typically be
somewhere between 20\% and 80\% of the on-disk size of the \texttt{SSW} file
from which it was converted.

Finally it is possible, via the \texttt{-c~FILE} flag, to point the
\texttt{ssw2mcpl} command to the input deck file used when producing the
provided \texttt{SSW} file. Doing so will result in a complete copy of that file
being stored in the \texttt{MCPL} header as a binary data blob under the string
key \texttt{"mcnp\_input\_deck"}, thus providing users with a convenient
snapshot in the \texttt{MCPL} file of the \texttt{MCNP} setup used. Unfortunately,
it was not possible to automate this procedure completely, and it thus relies
on the user to provide the correct input deck for a given \texttt{SSW} file. But
the \texttt{ssw2mcpl} command does at least check that the specified file is a
text-file and that it contains somewhere the correct value of the so-called \emph{problem title}: a
custom free-form string which is specified by the user in the input deck and embedded in the
\texttt{SSW} file by \texttt{MCNP}. The input deck embedded in a given \texttt{MCPL} file can
later be inspected from the command line by invoking the command
``\texttt{mcpltool~-bmcnp\_input\_deck~<file.mcpl>}''.

Usage of the \texttt{mcpl2ssw} command, for which the full usage instructions
are shown in Listing~\ref{lst:mcpl2sswusage}, is slightly more involved: in
addition to an input \texttt{MCPL} file, the user must also supply a reference
\texttt{SSW} file in a format suitable for the \texttt{MCNP} setup in which the
resulting \texttt{SSW} file is subsequently intended to be used as input. The
need for this added complexity stems from the constraint that the \texttt{SSW}
format is merely intended as an internal format in which it is possible to stop
and restart particles while remaining within a given setup of an \texttt{MCNP}
simulation -- meaning at the very least that the \texttt{MCNP} version and the
configuration of the geometrical surfaces involved in the \texttt{Surface Source
  Write/Read} procedure must be unchanged. Thus, for maximal robustness, the
user must supply a reference \texttt{SSW} file which was produced by the setup
in which the \texttt{SSW} file created with \texttt{mcpl2ssw} is to be used (it
does not matter how many particles the reference file contains). What will
actually happen is that in addition to the particle state data itself, the
newly created \texttt{SSW} file will contain the exact same header as the one in
the reference \texttt{SSW} file, apart from the fields related to the number of
particles in the file.

\begin{lstlisting}[float,language={},
  basicstyle={\linespread{0.9}\ttfamily\scriptsize},
  label={lst:mcpl2sswusage},
  caption={Usage instructions for the \texttt{mcpl2ssw} command.}
]
Usage:

  mcpl2ssw [options] <input.mcpl> <reference.ssw> [output.ssw]

Converts the Monte Carlo particles in the input MCPL file to SSW format
(MCNP Surface Source Write) and stores the result in the designated output
file (defaults to "output.ssw").

In order to do so and get the details of the SSW format correct, the user
must also provide a reference SSW file from the same approximate setup
(MCNP version, input deck...) where the new SSW file is to be used. The
reference SSW file can of course be very small, as only the file header is
important (the new file essentially gets a copy of the header found in the
reference file, except for certain fields related to number of particles
whose values are changed).

Finally, one must pay attention to the Surface ID assigned to the
particles in the resulting SSW file: Either the user specifies a global
one with -s<ID>, or it is assumed that the MCPL userflags field in the
input file is actually intended to become the Surface ID. Note that not
all MCPL files have userflag fields and that valid Surface IDs are
integers in the range 1-999999.

Options:

  -h, --help   : Show this usage information.
  -s<ID>       : All particles in the SSW file will get this surface ID.
  -l<LIMIT>    : Limit the number of particles transferred to the SSW file
                 (defaults to 2147483647, the maximal SSW capacity).
\end{lstlisting}

Additionally, the user must consider carefully which \texttt{MCNP} surface IDs
the particles from the \texttt{MCPL} file should be associated with, once
transferred to the \texttt{SSW} file. By default it will assume that the
\texttt{MCPL} user-flags field contains exactly this ID, but more often than
not, users will have to specify a global surface ID for all of the particles
through the \texttt{-s<ID>} command-line option for the \texttt{mcpl2ssw}
command.

Finally, note that \texttt{SSW} files do not contain polarisation information,
and any such polarisation information in the input \texttt{MCPL} file will
consequently be discarded in the translation. Likewise, in cases where the input
\texttt{MCPL} file contains one or more particles whose type does not have a
representation in the targeted flavour of \texttt{MCNP}, they will be ignored
with suitable warnings.

\subsection{\texttt{McStas} and \texttt{McXtrace} interfaces}\label{sec:plugins_mcstasmcxtrace}

Recent releases of the neutron ray tracing software package
\texttt{McStas}~\cite{mcstas1,mcstas2} (version 2.3 and later) and its X-ray
sibling package \texttt{McXtrace}~\cite{mcxtrace1} (version 1.4 and later)
include \texttt{MCPL}-interfaces. Although \texttt{McStas} and \texttt{McXtrace}
are two distinct software packages, they are implemented upon a common
technological platform, \texttt{McCode}, and the discussions here will for
simplicity use the term \texttt{McCode} where the instructions are otherwise
identical for users of the two packages.

The particle model adopted in \texttt{McCode} is directly compatible with
\texttt{MCPL}. In essence, apart from simple unit conversions, particles are read
from or written to \texttt{MCPL} files at one or more predefined logical points
defined in the \texttt{McCode} configuration files (so-called \emph{instrument
  files}). Specifically, two new components, \texttt{MCPL\_input} and
\texttt{MCPL\_output}, are provided, which users can activate by adding
entries at relevant points in their instrument files as is usual when working
with \texttt{McCode}.

First, when using the \texttt{MCPL\_input} component, particles are directly
read from an \texttt{MCPL} input file and injected into the simulation at the
desired point, thus playing the role of a source. In
Listing~\ref{lst:mccode_mcplinput1} is shown how, in its simplest form, users
would insert an \texttt{MCPL\_input} component in their instrument file. This
will result in the \texttt{MCPL} file being read in its entirety, and all found
neutrons (for \texttt{McStas}) or gamma particles (for \texttt{McXtrace}) traced
through the \texttt{McCode} simulation.  Listing~\ref{lst:mccode_mcplinput2}
indicates how the user can additionally impose an allowed energy range when
loading particles by supplying the \texttt{Emin} and \texttt{Emax}
parameters. The units are \si{meV} and \si{keV} respectively for \texttt{McStas}
and \texttt{McXtrace}. Thus, the code in Listing~\ref{lst:mccode_mcplinput2}
would select \SIrange{12}{100}{meV} neutrons in \texttt{McStas} and
\SIrange{12}{100}{keV} gammas in \texttt{McXtrace}.  A particle from the
\texttt{MCPL} file is injected at the position indicated by its \texttt{MCPL}
coordinates \emph{relative} to the position of the \texttt{MCPL\_input}
component in the \texttt{McCode} instrument. Thus, a user can impose coordinate
transformations by altering the positioning of \texttt{MCPL\_input} as shown in
Listing~\ref{lst:mccode_mcplinput3}, which would shift the initial position of
the particles by $(X,Y,Z)$ and rotate their initial velocities around the $x$,
$y$ and $z$ axes (in that order) by respectively $Rx$, $Ry$ and $Rz$
degrees. Furthermore, Listing~\ref{lst:mccode_mcplinput3} shows a way to
introduce a time shift of \SI{2}{s} to all particles, using an \texttt{EXTEND}
code block.

\begin{lstlisting}[float,language={[mccode]C},
  label={lst:mccode_mcplinput1},
  caption={Code enabling \texttt{MCPL} input in its simplest form.}
]
COMPONENT vin = MCPL_input( filename="myfile.mcpl" )
AT(0,0,0) RELATIVE Origin
\end{lstlisting}

\begin{lstlisting}[float,language={[mccode]C},
  label={lst:mccode_mcplinput2},
  caption={Code enabling \texttt{MCPL} input, selecting particles in a given energy range.}
]
COMPONENT vin = MCPL_input( filename="myfile.mcpl",
                            Emin=12, Emax=100 )
AT(0,0,0) RELATIVE Origin
\end{lstlisting}

\begin{lstlisting}[float,language={[mccode]C},
  label={lst:mccode_mcplinput3},
  caption={Code enabling \texttt{MCPL} input, applying spatial and temporal transformations.}
]
COMPONENT vin = MCPL_input( filename="myfile.mcpl" )
AT(X,Y,Z) RELATIVE Origin
ROTATED (Rx,Ry,Rz) RELATIVE Origin
EXTEND
%{
        t=t+2;
%}
\end{lstlisting}

For technical reasons, the number of particles to be simulated in
\texttt{McCode} must be fixed at initialisation time. Thus, the number of
particles will be set to the total number of particles in the input file, as
this is provided through the corresponding \texttt{MCPL} header field. If and
when a particle is encountered which can not be used (due to having a wrong
particle type or energy), it will lead to an empty event in which no particles
leave the source. At the end of the run, the number of particles skipped over
will be summarised for the user. This approach obviates the need for running
twice over the input file and avoids the potential introduction of
statistical bias from reading a partial file.

Note that if running \texttt{McCode} in parallel processing mode using
MPI~\cite{mpi_standard_2015}, each process will operate on all particles in the
entire file, but the particles will get their statistical weights reduced
accordingly upon load. This behaviour is not specific to the
\texttt{MCPL\_input} component, but is a general feature of how multiprocessing
is implemented in \texttt{McCode}.

When adding an \texttt{MCPL\_output} component to a \texttt{McCode} instrument
file, the complete state of all particles reaching that component is written to
the requested output file. In Listing~\ref{lst:mccode_mcploutput1} is shown how,
in its simplest form, users would insert such a component in their instrument
file, and get particles written with coordinates relative to the component
preceding it, into the output file (replace \texttt{RELATIVE PREVIOUS} with
\texttt{RELATIVE ABSOLUTE} to write absolute coordinates instead). For reference, a copy of the complete
instrument file is stored in the \texttt{MCPL} header as a binary data blob
under the string key \texttt{"mccode\_instr\_file"}. This feature provides users
with a convenient snapshot of the generating setup. The instrument file embedded
in a given \texttt{MCPL} file can be inspected from the command line by invoking
the command ``\texttt{mcpltool -bmccode\_instr\_file <file.mcpl>}''.

If running \texttt{McCode} in parallel processing mode using MPI, each process
will create a separate output file named after the pattern
\texttt{myoutput.node\_idx.mcpl} where \texttt{idx} is the process number (assuming
\texttt{filename="myoutput.mcpl"} as in
Listing~\ref{lst:mccode_mcploutput1}), and those files will be automatically
merged during post-processing into a single file.\footnote{This automatic merging
only happens in \texttt{McStas} version 2.4 or later and  \texttt{McXtrace}
version 1.4 or later, and can be disabled by setting the parameter \texttt{merge\_mpi=0}. Users of earlier versions will have to use the
\texttt{mcpltool} command to perform the merging manually, if desired.}

\begin{lstlisting}[float,language={[mccode]C},
  label={lst:mccode_mcploutput1},
  caption={Code enabling \texttt{MCPL} output in its simplest form.}
]
COMPONENT mcplout = MCPL_output( filename="myoutput.mcpl" )
AT(0,0,0) RELATIVE PREVIOUS
\end{lstlisting}

To avoid generating unnecessarily large files, the \texttt{MCPL\_output}
component stores particle state data using the global PDG code feature
(cf.\ section~\ref{sec:format_infoavail}), uses single-precision floating point
numbers, and does \emph{not} by default store polarisation vectors. The two
latter settings may be changed by the user through the \texttt{polarisationuse}
and \texttt{doubleprec} parameters respectively, as shown in
listing~\ref{lst:mccode_mcploutput2}.

\begin{lstlisting}[float,language={[mccode]C},
  label={lst:mccode_mcploutput2},
  caption={Code enabling \texttt{MCPL} output with polarisation and double-precision numbers.}
]
COMPONENT mcplout = MCPL_output( filename="myoutput.mcpl",
                                 polarisationuse=1,
                                 doubleprec=1 )
AT(0,0,0) RELATIVE PREVIOUS
\end{lstlisting}

Finally, if desired, custom information might be stored per-particle into the
\texttt{MCPL} user-flags field for later reference. This could be any property,
such as for instance the number of reflections along a neutron guide, or the
type of scattering process in a crystal,
etc. Listing~\ref{lst:mccode_mcploutput3} shows a simple example of this where
the particle ID, in the form of its \texttt{McCode} ray number (returned from
the \texttt{McCode} library function \texttt{mcget\_run\_num}), is stored into
the user-flags field. A string, \texttt{userflagcomment}, is required in order
to describe the significance of the extra data, and will end up as a comment in
the resulting \texttt{MCPL} file.

\begin{lstlisting}[float,language={[mccode]C},
  label={lst:mccode_mcploutput3},
  caption={Code enabling \texttt{MCPL} output with custom user-flags information.}
]
/* some upstream component setting a variable (customvar) */
COMPONENT some_comp = Some_Component( /* some parameters here */ )
AT (0,0,0) ABSOLUTE
EXTEND %{
  customvar=(uint32_t) mcget_run_num();
%}

/* MCPL output capturing customvar into MCPL user-flags */
COMPONENT vout = MCPL_output( filename="myoutput.mcpl",
                              userflag=customvar,
                              userflagcomment="Particle Id" )
AT(0,0,0) RELATIVE PREVIOUS
\end{lstlisting}

\section{Example scientific use cases}\label{sec:examples}

The possible uses for \texttt{MCPL} are envisioned to be many and varied,
facilitating both straight-forward transfers of particle data between different
simulations, as well as data reuse and cross-code comparisons. Actual scientific
studies are already being performed with the help of \texttt{MCPL},
demonstrating the suitability of the format ``in the field''.  By way of
example, it will be discussed in the following how \texttt{MCPL} is used in two
such ongoing studies.

\subsection{Optimising the detectors for the LoKI instrument at ESS}\label{sec:example_loki}

The ongoing construction of the European Spallation Source
(ESS)~\cite{esscdr,esstdr} has initiated significant development of novel
neutronic technologies in the past 5 years. The performance requirements for
neutron instruments at the ESS, in particular those resulting from the
unprecedented cold and thermal neutron brightness, are at or beyond the
capabilities of detector technologies currently available~\cite{kirstein2014}.
Additionally, shortage of $^3$He~\cite{he3crisis1,he3crisis2}, upon which the
vast majority of previous detectors were based, augments the need for
development of new efficient and cost-effective detectors based on other
isotopes with high neutronic conversion cross sections.

A typical approach to instrument design and optimisation at ESS involves the
development of a \texttt{McStas}-based simulation of the instrument. Such a
simulation includes an appropriate neutron source description and detailed
models of the major instrument components, such as benders, neutron guides,
chopper systems, collimators, sample environment and sample. See~\cite{carlile}
for an introduction to the role of the various instrument components. Detector
components in \texttt{McStas} are, however, typically not implemented with any
detailed modelling, and are simply registering all neutrons as they
arrive. Thus, while the setup in \texttt{McStas} allows for an efficient and
precise optimisation of most of the instrument parameters, detailed detector
optimisation studies must out of necessity be carried out in a separate
simulation package, such as \texttt{Geant4}.

As the detector development progresses in parallel with the general instrument
design, it is crucial to be able to optimise the detector setup for the exact
instrument conditions under investigation in \texttt{McStas}. The \texttt{MCPL}
format, along with the interfaces discussed in sections~\ref{sec:plugins_geant4}
and \ref{sec:plugins_mcstasmcxtrace}, facilitates this by allowing for easy
transfer of neutron states from the \texttt{McStas} instrument simulation into
\texttt{Geant4} simulations with detailed setups of proposed detector designs.

Technically, this is done by placing the \texttt{MCPL\_output} component just
after the relevant sample component in the \texttt{McStas} instrument
file. Additionally, using the procedure for creation and storage of custom
\texttt{MCPL} user-flags also discussed in
section~\ref{sec:plugins_mcstasmcxtrace}, it is possible to differentiate
neutrons that scattered on the sample from those which continued undisturbed, and
to carry this information into the \texttt{Geant4} simulations. This information
is needed to understand the impact of the direct beam on the low angle
measurements, in order to study the requirements for a so-called zero-angle
detector.

\begin{figure}
  \centering
  \includegraphics[scale=0.4]{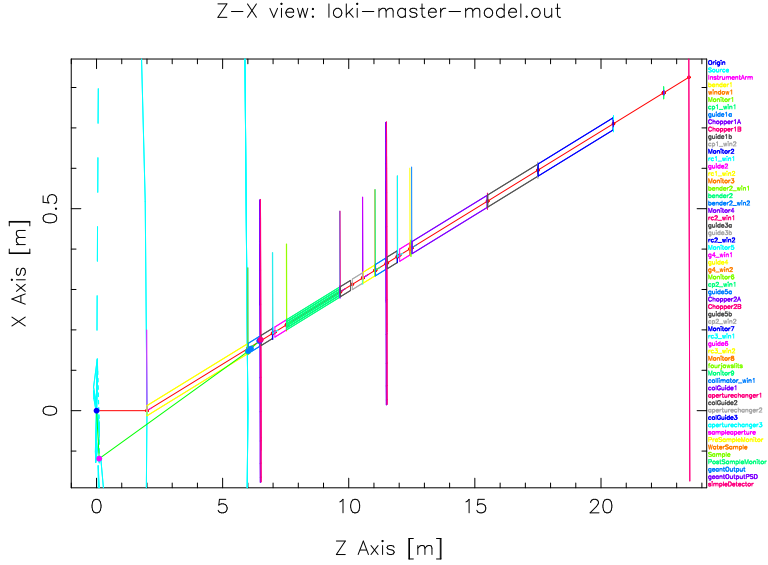}
  \caption{Layout of the \texttt{McStas} model of the LoKI instrument. Neutrons
    originate at the source located at $z=0$ and progress through the various
    instrument components toward the sample at $z=\SI{22.5}{m}$.}
  \label{loki_mcstas}
\end{figure}

\begin{figure}
  \centering
  \includegraphics[scale=0.4]{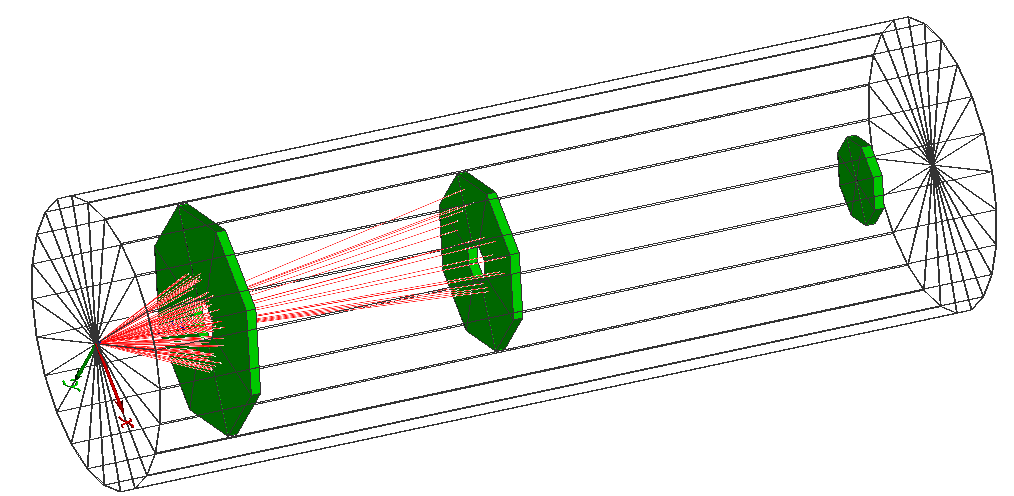}
  \caption{\texttt{Geant4} model of a potential detector geometry for the LoKI
    instrument. Neutrons from the sample hitting the active detector area appear in red.}
  \label{loki_g4}
\end{figure}

For example, in order to optimise the detector technology that the LoKI
instrument~\cite{lokijackson2015, lokikanaki2013, lokikanaki2013corr} might adopt, a series of \texttt{McStas}
simulations of the instrument components and the interactions in realistic
samples~\cite{sasviewkernels} are performed (see Figure~\ref{loki_mcstas} for a
view of the instrument in \texttt{McStas}). The parameters of the instrument and
the samples in the \texttt{McStas} model are chosen in such a way, that various
aspects of the detector performance can be investigated, including rate
capability and spatial resolution. The neutrons emerging from the sample in
\texttt{McStas} are then transferred via \texttt{MCPL} to the detector
simulation in \texttt{Geant4}, where a detailed detector geometry and
appropriate materials are implemented (see Figure~\ref{loki_g4} for a
visualisation of the \texttt{Geant4} model).

Neutrons traversing the detector geometry in \texttt{Geant4} undergo
interactions with the materials they pass on their flight-path, according to the
physics processes and respective cross sections available in the setup. Special
attention is needed when configuring the \texttt{Geant4} physics modelling, to
ensure that all processes relevant for neutron detection are taken into account
and handled correctly. Specifically, the setup utilises the
high-precision neutron models in \texttt{Geant4} extended with~\cite{nxsg4}, and
is implemented in~\cite{dgcodechep2013}. In the solid-converter based detectors
under consideration, a neutron absorption results in emission of charged products which then
travel a certain range inside the detector and deposit energy in a counting
gas. It is possible to extract position and time information from the energy
deposition profile and use these space-time coordinates for further analysis, in
the same way that measurements in a real detector would be treated. This way it becomes
possible to reproduce the distributions of observable quantities relevant for
Small Angle Neutron Scattering (SANS) analysis~\cite{sansfeigin, sansimae}.

\begin{figure}
  \centering
  \includegraphics[scale=0.5]{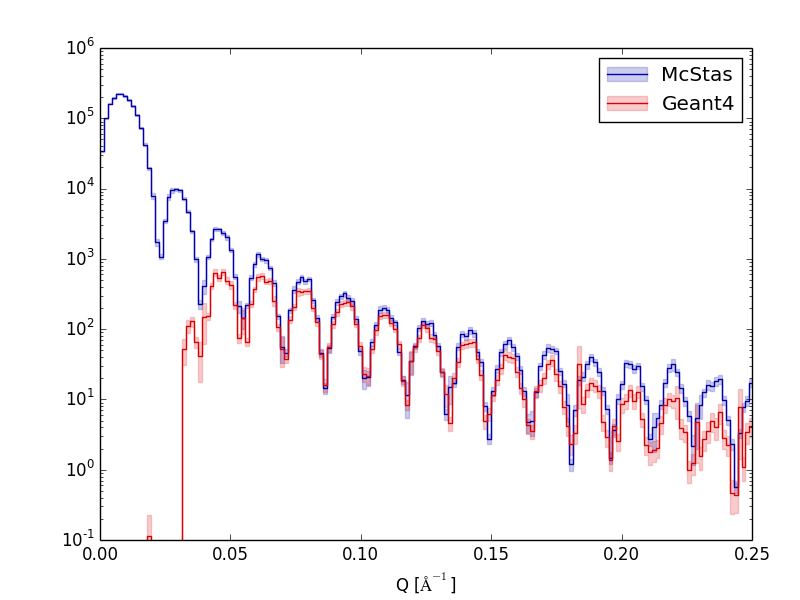}
  \caption{Raw $Q$ distribution for a subset of the LoKI detectors
    (middle detector bank of Figure~\ref{loki_g4}). The
    \texttt{McStas} post-sample output appears in blue, while the distribution calculated
    from the simulated measurements in \texttt{Geant4} appears in red.}
  \label{loki_q}
\end{figure}

One such observable quantity is the $Q$ distribution~\cite[Ch.~2.3.3]{carlile},
where $Q$ is defined as the momentum change of the neutron as it scatters on the
sample, divided by $\hbar$: $Q\equiv|\Delta\vec{p}|/\hbar$. Figure~\ref{loki_q}
demonstrates such a distribution, based on the simulated
output of the middle detector bank of LoKI (cf.\ Figure~\ref{loki_g4}), for a
certain instrument setup -- including a sample modelled as consisting of spheres
with radii of \SI{200}{\angstrom}.  The raw $Q$ distribution is calculated both
based on the neutron states as they emerge from the sample in \texttt{McStas},
and from the simulated measurements in \texttt{Geant4}.  With such a procedure,
resolution-smearing effects can be correctly attributed to their sources,
geometrical acceptance and detector efficiency can be studied in detail, and the
impact of engineering features such as dead space can be accurately considered.

\subsection{Neutron spectra predictions for cosmogenic dating studies}

The use of radionuclides produced in-situ by cosmic rays for dating purposes
has, in the last two decades, revolutionised the earth surface
sciences~\cite{Dunai2010}. The precise determination of the production rate of
such isotopes, like $^{10}$Be and $^{26}$Al, poses the key challenge for this
technique and relies on a folding of cosmic fluxes with energy dependent
production cross sections~\cite{Reedy2013}. The present discussion will focus on
the evaluation of the neutron flux induced by cosmic radiation, and in
particular on how \texttt{MCPL} can be exploited both to facilitate the reuse of
computationally intensive simulations, and as a means for cross-code
comparisons.

At sea level, neutrons constitute the most abundant hadronic component of cosmic
ray induced showers, and possess relatively high cross sections for production
of isotopes relevant for radionuclide dating. Thus, it is the dominant
contributor to the relevant isotopic production in the first few meters below
the surface~\cite{Gosse2001}.  Extending further below the surface, the neutron
flux decreases rapidly, and as a consequence the isotopic production rate
induced by cosmic muons eventually becomes the most significant
factor~\cite{Heisinger2002a,Heisinger2002b}.  At a depth of approximately
\SI{3}{\meter} below the surface, the production rate due to muons is comparable
with the rate from neutrons~\cite{Gosse2001}. Considering non-erosive surfaces
and samples at depths significantly less than \SI{3}{\meter}, the production
rates can thus be estimated by considering just the flux of neutrons. Thus,
given known cross sections for neutronic production of $^{10}$Be or $^{26}$Al,
properties such as the cosmic irradiation time of a given sample can be directly
inferred from its isotopic content -- providing information about geological
activity. In the present study, Monte Carlo methods are used to simulate
atmospheric cosmic rays~\cite{Masarik1999,Masarik2009} and subsequently estimate
the neutron flux spectra as a function of depth under the surface of the Earth.

\begin{figure}
  \centering
  \includegraphics[width=.6\textwidth]{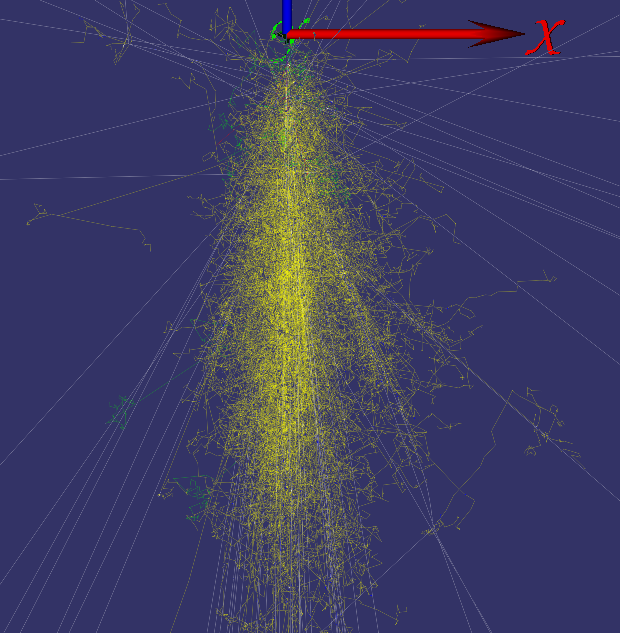}
  \caption{Cosmic shower simulated in \texttt{Geant4}. The incident proton
    energy is \SI{100}{\GeV} and the length of the $x$-axis is \SI{2}{\km}. The
    straight grey trajectories are neutrinos. The yellow and green trajectories
    are photons and neutrons, respectively.  }
  \label{fShower}
\end{figure}

Primary cosmic rays constantly bombard the solar system and initiate cosmic ray
showers in the Earth's atmosphere, leading to the production of atmospheric
neutrons. Figure~\ref{fShower} shows the trajectories of a simulated air shower
induced by a single \SI{100}{\GeV} proton in \texttt{Geant4}: very large numbers
of secondary particles are generated in each shower, all of which must
themselves undergo simulation. Full scale simulation of such showers is
therefore relatively time consuming.  On the other hand, simulations of the
propagation of sea level neutrons in a few meters of solid material are
relatively fast.  In the present work of estimating neutron spectra for
different underground materials, \texttt{MCPL} is used to record particle
information at sea level. Using the recorded data as input, subsequent
simulations are dedicated to the neutron transport in different underground
materials. In this way, repetition of the time consuming parts of the simulation
is avoided.  \texttt{Geant4} is used to simulate the air shower in this work,
while both \texttt{Geant4} and \texttt{MCNPX} are used to simulate neutron
spectra underground.

In the \texttt{Geant4} simulation of the Earth's atmosphere, the geometry is
implemented as a \SI{100}{km} thick shell with an inner radius of \SI{6387}{km},
sub-divided into 50 equally thick layers, the effective temperatures and
densities of which are calculated using the ``U.S.  standard atmosphere, 1976''
model~\cite{Jursa1985}.  Using the plugins described in
section~\ref{sec:plugins_geant4}, the simulation of any particle reaching the
inner surface of the atmosphere is ended and its state stored in an
\texttt{MCPL} file.  To compare the simulated and measured~\cite{Gordon2004}
spectra at New York city, a lower cutoff of $E_c=\SI{2.08}{GeV}$ on the kinetic
energy of the primary proton is applied, to take the geomagnetic field shielding
effect at this location into account.  The relationship between the number of
simulated primary protons, $N$, and the real world time-span, ${\delta}t$, to
which such a sample-size corresponds, is given by the following equation:
\begin{equation*}
{\delta}t=\frac{N}{\int\limits_{E_{c}}^{\infty} J(E) dE \times 2\pi\times 4\pi r^2 }
\end{equation*}
Here, $r$ is the outer radius of the simulated atmosphere and $J$ the
differential spectrum of Usoskin's model~\cite{Usoskin2005} using the
parameterisation in~\cite{Herbst2010}.

In the simulation of \SI{4.20e6} primary protons, the resulting integral
neutronic flux above \SI{20}{\MeV} at sea level was found to be
\SI{3.27e-15}{\per\square\cm}, corresponding to an absolute surface flux at New
York city of \SI{4.22e-3}{\per\square\cm\per\second}.  Integrating the measured
reference neutron spectrum tabulated in~\cite{Gordon2004} above \SI{20}{\MeV},
an integral flux of \SI{3.15e-3}{\per\square\cm\per\second} is obtained.  The
simulation thus overestimates the measured flux by 34\%, which is a level of
disagreement compatible with the variation in the predicted value of the integral flux
between different models of the local interstellar spectrum~\cite{Herbst2010}.  Therefore,
the performance of the atmospheric simulation is concluded to be satisfactory.

\begin{figure}
  \centering
  \includegraphics[width=.8\textwidth]{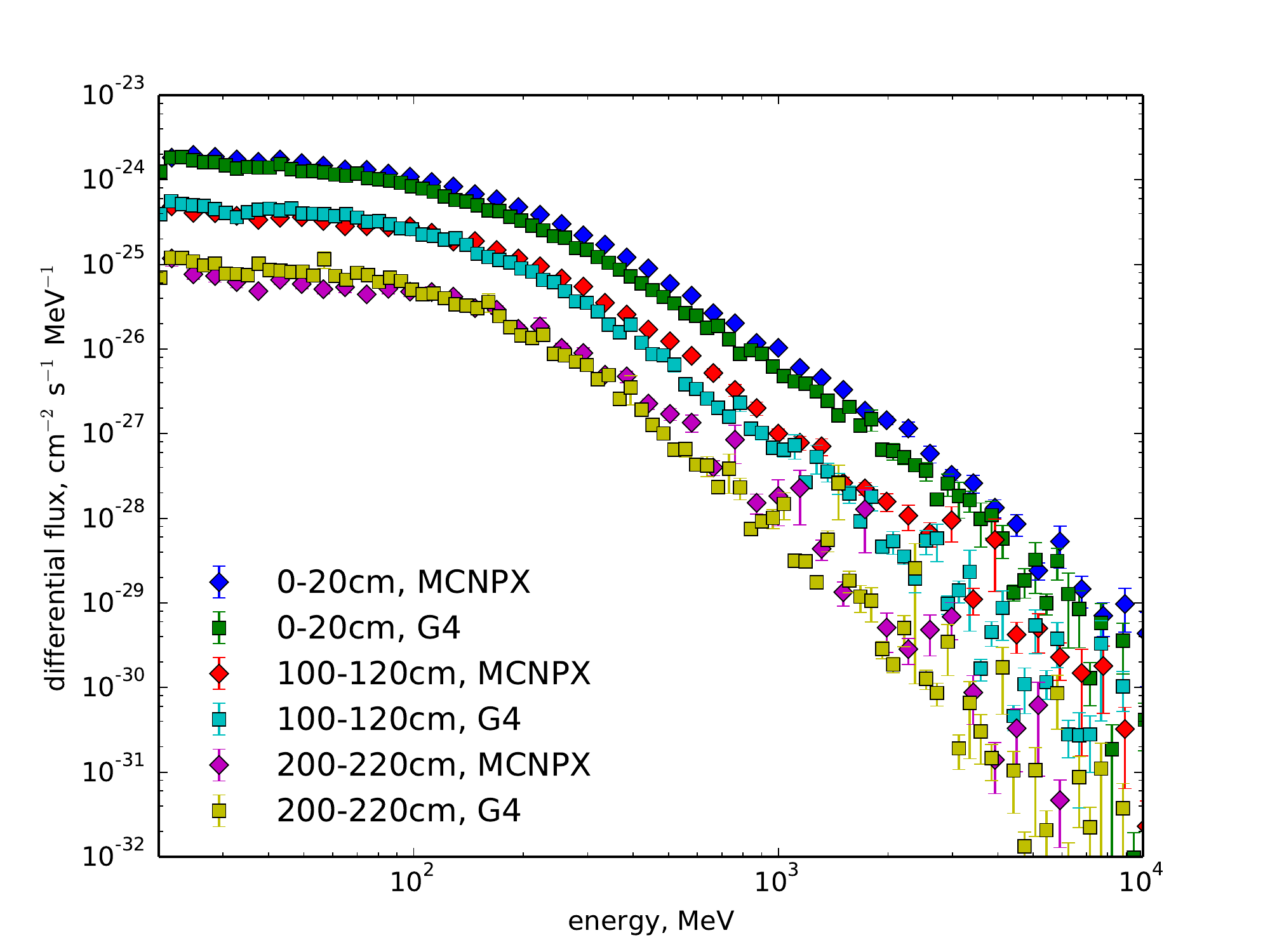}
  \caption{Comparisons of simulated neutron spectra in underground quartz.}
  \label{fQuartz}
\end{figure}

In the subsequent underground simulations presented here, the Earth is for
simplicity modelled as consisting entirely of quartz (SiO$_2$), which is a
sample material widely used in cosmogenic dating applications~\cite{Dunai2010},
as both $^{26}$Al and $^{10}$Be are produced within when subjected to neutron
radiation -- normally via spallation.  The \texttt{MCPL} files generated by the
computationally expensive atmospheric shower simulation described above, is
input to the underground simulations implemented in both \texttt{Geant4} and
\texttt{MCNPX}, using the interfaces described in
sections~\ref{sec:plugins_geant4} and \ref{sec:plugins_mcnp}. The geometries in
both cases are defined as \SI{20}{\cm} thick spherical shells consisting of pure
quartz.  As the threshold energies of the related spallation reactions are well
above \SI{20}\MeV, only spectra above this energy are compared in this study.
The simulated volume spectra in a few layers are compared in
Figure~\ref{fQuartz}. Good agreement between \texttt{Geant4} and \texttt{MCNPX}
is observed.

In conclusion, a useful method for disentangling the resource intensive
simulation of cosmic showers from subsequent faster simulations of neutron
transport in the Earth crust has been demonstrated using \texttt{MCPL} as an
intermediate stepping stone.  The simulation strategy thus employed eases the
use of computational resources, and provides a means for cross-comparison
between simulation codes.  Given reliable energy dependent cross sections, many
of the key parameters for cosmogenic dating applications can be provided based
on the work described in this section.

\section{Summary and outlook}

The \texttt{MCPL} format provides flexible yet efficient storage of
particle-state information, aimed at simplifying and standardising interchange
of such data between applications and processes. The core parts of \texttt{MCPL}
are implemented in portable and legally unencumbered \texttt{C} code. This is
intended to facilitate adoption into existing packages and build systems, and
the creation of application-specific converters and plugins.

In connection with the initial release presented here, \texttt{MCPL} interfaces
were created for several popular Monte Carlo particle simulation packages:
\texttt{Geant4}, \texttt{MCNP}, \texttt{McStas} and \texttt{McXtrace}. It is the
intention and hope that the number of such \texttt{MCPL}-aware applications will
increase going forward. A website~\cite{mcplwww} has been set up for the
\texttt{MCPL} project, on which users will be able to locate future updates to
the \texttt{MCPL} distribution, as well as relevant documentation.

\section*{Acknowledgements}

This work was supported in part by the European Union's Horizon 2020 research
and innovation programme under grant agreement No 676548 (the BrightnESS
project) and under grant agreement No 654000 (the SINE2020 project).

\section*{References}

\bibliography{refs_thki}

\vfill\clearpage

\appendix

\section{Detailed layout of \texttt{MCPL} files}\label{appendix:mcpldetailedlayout}

It is recommended to manipulate \texttt{MCPL} files through (direct or indirect)
calls to functions in \texttt{mcpl.h}. This not only ensures consistency and
simplicity, but also allows painless format evolution where clients simply need
to obtain updated versions of \texttt{mcpl.h} and \texttt{mcpl.c} in order to
support new versions of the \texttt{MCPL} file format, while typically retaining
full backwards compatibility. Nonetheless, this appendix provides reference
information concerning the exact binary layout of data in \texttt{MCPL} files,
specifically those in the third version of the format (written by \texttt{MCPL}
starting from version 1.1.0). The file format version of a given \texttt{MCPL}
file is encoded in the first few bytes of a file (see below), is printed upon
inspection with the \texttt{mcpltool} command line utility, and is available
programmatically with the function \fctname{mcpl_hdr_version} and macro
\fctname{MCPL_FORMATVERSION} (cf.~\ref{appendix:reference_c_api}). Note that all
floating point numbers in \texttt{MCPL} files must adhere to the
\texttt{IEEE-754} floating point standard~\cite{IEEE754} for single- (32 bit) or double-precision (64 bit) as relevant. All signed integers
in the file must follow the ubiquitous ``two's complement'' representation.

The first 48 bytes of an \texttt{MCPL} file follow a fixed layout, as indicated
in Table~\ref{tab:appendix_hdr}, providing flags and values needed to read and
interpret both the particle data section and the remainder of the header
section. Note that abbreviations used for type information in tables in this
appendix are UINT32 and UINT64 for 32 and 64 bit unsigned integers respectively,
INT32 for 32 bit signed integers and FP64 for double-precision 64 bit floating point
numbers.

\begin{table}[tbp]
\centering
\resizebox{\textwidth}{!}{%
\begin{tabular}{@{}ccccccp{0.8\textwidth}@{}}
\toprule
\multicolumn{7}{c}{\textbf{Header layout (first part)}} \\ \midrule
\textit{Position} && \textit{Size} && \textit{Type} && \textit{Description} \\ \cmidrule(){1-1} \cmidrule(){3-3} \cmidrule(){5-5} \cmidrule(){7-7}
0 && 4B && - && Magic number identifying file as an \texttt{MCPL} file. Value is always 0x4d43504c ("MCPL" in \texttt{ASCII}).\\
4 && 3B && - && File format version encoded as 3 digit zero-prefixed \texttt{ASCII} number (e.g. "003").\\
7 && 1B && - && Endianness of numbers in file, 0x4c (\texttt{ASCII} ``L'') for \emph{little} or 0x42 (\texttt{ASCII} ``B'') for \emph{big}.\\
8 && 8B && UINT64 && Number of particles in file.\\
16 && 4B && UINT32 && Number of custom comments in file, \textbf{NCMTS}.\\
20 && 4B && UINT32 && Number of custom binary data blobs in file, \textbf{NBLOBS}.\\
24 && 4B && UINT32 && Flag signifying whether user-flags are enabled (0x1) or disabled (0x0).\\
28 && 4B && UINT32 && Flag signifying whether polarisation vectors are enabled (0x1) or disabled (0x0).\\
32 && 4B && UINT32 && Flag signifying whether floating point numbers in the particle data section are in double- (0x0) or single-precision (0x1).\\
36 && 4B &&  INT32 && Value of universal PDG code. A value of 0x0 means that particles in the file all have their own PDG code field.\\
40 && 4B && UINT32 && Data length per particle (redundant information, as it can be inferred from the other flags and values).\\
44 && 4B && UINT32 && Flag signifying whether a universal weight is present in the header (0x1) or if particles in the file all have their own weight field (0x0).\\
\bottomrule
\end{tabular}%
}
\caption{Detailed layout of the first part of the header section of an \texttt{MCPL} file.}
\label{tab:appendix_hdr}
\end{table}

The layout of the second part of the header section is indicated in
Table~\ref{tab:appendix_hdr2}, and includes both optional and repeated entries
and entries with flexible length. Entries with type listed as ``DATA ARRAY'' are
arbitrary length byte-arrays in which the first four bytes are unsigned 32 bit
integers indicating the byte length of the data payload which follows just
after. Note that strings are stored like any other data, with the only twist
being that terminating \texttt{NULL} characters are \emph{not} stored.

\begin{table}[tbp]
\centering
\resizebox{\textwidth}{!}{%
\begin{tabular}{@{}ccccp{0.8\textwidth}@{}}
\toprule
\multicolumn{5}{c}{\textbf{Header layout (second part)}} \\ \midrule
\textit{Size} && \textit{Type} && \textit{Description} \\ \cmidrule(){1-1} \cmidrule(){3-3} \cmidrule(){5-5}
0B or 8B && FP64 && Value of the universal weight, if enabled.\\
4B+ && DATA ARRAY && Data is a string holding the user provided ``Source name''. This is always present, but might be empty.\\
\textbf{NCMTS} $\times$ 4B+ && DATA ARRAYS && One data array for each comment, holding the user provided comments as strings.\\
\textbf{NBLOBS} $\times$ 4B+ && DATA ARRAYS && One data array for each binary data blob, holding the user provided blob keys as strings.\\
\textbf{NBLOBS} $\times$ 4B+ && DATA ARRAYS && One data array for each binary data blob, holding the actual binary data (in the same order as the blob keys).\\
\bottomrule
\end{tabular}%
}
\caption{Detailed layout of the second part of the header section of an \texttt{MCPL} file. The presence and count of entries here depends on values found in the first part of the header section (cf.~Table~\ref{tab:appendix_hdr}).}
\label{tab:appendix_hdr2}
\end{table}

Next, after the header section, the remainder of the file consists of the
particle data section. It contains one entry for each particle in the file, with
detailed layout of each as indicated in Table~\ref{tab:appendix_pdata}. Here, FP
is either single- (32 bit) or double-precision (64 bit) numbers, depending on the
relevant flag in the file header. Concerning the 3 floating point numbers used
to represent the packed direction vector and kinetic energy, the scheme is as
discussed in \ref{appendix:unitvectorpacking}: the first and second of the 3
floating point numbers are respectively FP1 and FP2 from
Table~\ref{tab:adaptivepp}, while the third is a number whose magnitude is given
by the particle's kinetic energy and whose sign bit is used to store the last bit
of information needed for the direction vector, indicated in the ``+1 bit'' column of
Table~\ref{tab:adaptivepp}. Specifically, the sign bit is set when the number
indicated in the ``+1 bit'' column is negative.

\begin{table}[tbp]
\centering
\resizebox{\textwidth}{!}{%
\begin{tabular}{@{}ccccp{0.8\textwidth}@{}}
\toprule
\multicolumn{5}{c}{\textbf{Particle data layout}} \\ \midrule
\textit{Presence} && \textit{Count \& type} && \textit{Description} \\ \cmidrule(){1-1} \cmidrule(){3-3} \cmidrule(){5-5}
OPTIONAL && 3 $\times$ FP && Polarisation vector (if enabled in file).\\
ALWAYS && 3 $\times$ FP && Position vector \\
ALWAYS && 3 $\times$ FP && Packed direction vector and kinetic energy.\\
ALWAYS && 1 $\times$ FP && Time.\\
OPTIONAL && 1 $\times$ FP && Weight (if file does not have universal weight).\\
OPTIONAL && 1 $\times$ INT32 && PDG code (if file does not have universal PDG code).\\
OPTIONAL && 1 $\times$ UINT32 && User-flags (if enabled in file).\\
\bottomrule
\end{tabular}%
}
\caption{Detailed layout of the data associated with each particle in an \texttt{MCPL} file.}
\label{tab:appendix_pdata}
\end{table}

\vfill\clearpage

\section{Unit vector packing}\label{appendix:unitvectorpacking}

From a purely mathematical perspective, it is trivial to ``pack'' unit vectors
specified in three-dimensional Cartesian coordinates into a two-dimensional
coordinate space, and it can for instance be achieved by the standard
transformation between Cartesian and Spherical coordinates. However, when
considering floating point numbers rather than the ideal mathematical
abstraction of the complete set of real numbers, issues of numerical imprecision
during the packing and subsequent unpacking transformations become
crucial. Where it matters, the discussion in this appendix will, like the
\texttt{MCPL} format in general, assume floating point numbers to adhere to the
relevant \texttt{IEEE} floating point standard~\cite{IEEE754}.

A few different packing algorithms will be compared in this appendix. First of
all, the spherical transformation between Cartesian unit vectors $(u_x,u_y,u_z)$
and spherical coordinates $(\theta,\phi)$ is investigated, using the \texttt{C}
math library functions \texttt{acos}, \texttt{atan2}, \texttt{sin} and
\texttt{cos} in an obvious manner when implementing the transformations. In
addition to the significant computational overhead involved when evaluating
trigonometric functions, numerical uncertainties also tend to blow up when a
Cartesian coordinate is very small. For instance, consider a unit vector with
$u_z=\epsilon$ for some $|\epsilon|\ll1$. Then, $\theta=\arccos(\epsilon)$ which
is to lowest order equal to $\pi/2-\epsilon$, a subtraction which out of
necessity will cause a loss of precision when it is stored as a floating point
number: if $\epsilon$ is $N$ orders of magnitude smaller than $\pi/2$, then the
stored result will be insensitive to the $N$ least significant digits of
information in the storage of $\epsilon$. The calculation of $\phi$ suffers from
similar problems.

The next packing algorithm to be considered is what will be denoted the
\emph{Static Projection} method in the following. It represents the
straight-forward and widespread solution of storing two Cartesian components,
$u_x$ and $u_y$, directly and recovering the magnitude of the third by the
expression $|u_z|=\sqrt{1-u_x^2-u_y^2}$. This method, which incidentally is the
one used internally in the \texttt{SSW} files produced by \texttt{MCNP}
(cf.\ section~\ref{sec:plugins_mcnp}), requires a single bit of additional
storage to be available, in order to recover the sign of $u_z$ as well as its
magnitude. This requirement of an extra bit of storage is seen in several
packing schemes and is not necessarily a problem, as will be discussed
later. What is problematic, however, is that the calculation of $\sqrt{1-u_x^2-u_y^2}$
will result in large numerical uncertainties when the magnitude of $u_z$ is
small, as it implies the subtraction of large and nearly equal quantities and a
resulting loss of significant digits in the result.

Finally, inspired by a recent survey of unit vector packing
techniques~\cite{Cigolle2014Vector}, a packing scheme using a so-called
\emph{Octahedral Projection}~\cite{octenvmaps2008} was also investigated. It is
a variant of the Static Projection method in which the original point on the
unit sphere is first projected onto an octahedral surface before the resulting
$x$ and $y$ coordinates are stored. Thus, the stored variables are
$(o_x,o_y)=(u_x/n,u_y/n)$ where $n=|u_x|+|u_y|+|u_z|$. Unpacking proceeds by
first recovering the point on the octahedron as $(o_x,o_y,1-|o_x|-|o_y|)$, and
then projecting back out to the unit sphere with a simple normalisation,
requiring the evaluation of a square root. Although the algorithm has improved
performance over the methods already discussed, it once again suffers from
numerical precision issues when applied to certain unit vectors. Consider for
instance the unit vector $(\sqrt{1-\epsilon^2},0,\epsilon)$ for a very small but
positive epsilon. Packing will result in $o_y=0$ and
$o_x=1/(\sqrt{1-\epsilon^2}+\epsilon)\approx1-\epsilon$, storage of which
discards the $N$ least significant digits of $\epsilon$, when $\epsilon$ is $N$
orders of magnitude smaller than unity. Finally, it should be mentioned that
like the Static Projection method, the Octahedral Projection method also needs the
sign of $u_z$ stored in a bit elsewhere. The present discussion thus ignores the
method described in \cite{Cigolle2014Vector,octenvmaps2008} of encoding the sign
of $u_z$ by folding $(o_x,o_y)$ into a separate part of the plane when $u_z<0$,
as this operation introduces additional numerical imprecision and an undesired
asymmetry into the algorithm.

Although the three pre-existing packing methods discussed so far all represent
potentially useful approaches to unit vector packing depending on the particular
needs of a given use-case, they were nonetheless deemed undesirable for a
general purpose scientific format like \texttt{MCPL}. This is because
\texttt{MCPL} is intended for usage in a wide variety of simulation scenarios,
including those where some components of the directional unit vectors of stored
particles could be truly minuscule but non-zero in magnitude. In order to better
fulfil the requirements, a new unit vector packing algorithm was devised for
\texttt{MCPL}.  It is tentatively named \emph{Adaptive Projection Packing}, and
will be presented in the following. The algorithm is based upon the simple
observation that although the Static Projection method has issues when
$|u_z|\ll1$, it provides very precise results when $u_z$ is the component with
the largest magnitude. Thus, rather than always storing $u_x$ and $u_y$,
performance can be significantly improved by letting the packing algorithm
select the two components with the smallest magnitudes and store those. That
leaves the issue that the unpacking algorithm must be able to recognise which
components were stored. The solution to that is based upon the fact that all
three components can have at most unit magnitude, and that $|u_z|\le1/\sqrt2$
whenever it must be stored, implying that $1/u_z$ will have a magnitude larger
than $1$. Thus, by storing $1/u_z$ in place of the one of $u_x$ and $u_y$ which
is greater in magnitude, the unpacking algorithm can easily tell which
components are stored in which of the two packed numbers, merely by looking at
how their magnitudes compare to unity (the code should obviously store the
floating point representation of $\infty$ when $u_z=0$). The resulting encoding
scheme is summarised in Table~\ref{tab:adaptivepp}: one of three storage
scenarios is picked by the packing code, depending on the magnitudes of the
three unit vector components. The unique packed signature of each scenario,
listed in the last column, enables them to be easily distinguished by unpacking
code. Although obviously more complicated than the Static Projection method, the
added computational cost of using the Adaptive Projection method is at most a
few branches and a division, which is found in the context of \texttt{MCPL} to
be comparable to the cost of the Octahedral Projection method and much faster
than the Spherical coordinate method due to the expensive trigonometric function
calls. In any case, the overhead is found to be insignificant given the intended
usage in \texttt{MCPL}.

\begin{table}[tbp]
\centering
\begin{tabular}{@{}llclclcll@{}}
\toprule
\multicolumn{9}{c}{\textbf{Adaptive Projection Packing}} \\ \midrule
\textit{Scenario}         && \textit{FP1} && \textit{FP2} && \textit{$+$1 bit} && \textit{Packed signature}   \\%
 \cmidrule(){1-1} \cmidrule(){3-3}\cmidrule(){5-5} \cmidrule(){7-7} \cmidrule(){9-9}
$|u_x|$ largest && $1/u_z$ && $u_y$   && $\sign(u_x)$ && $|\text{FP1}|>1$, $|\text{FP2}|<1$  \\
$|u_y|$ largest && $u_x$   && $1/u_z$ && $\sign(u_y)$ && $|\text{FP1}|<1$, $|\text{FP2}|>1$ \\
$|u_z|$ largest && $u_x$   && $u_y$   && $\sign(u_z)$ && $|\text{FP1}|<1$, $|\text{FP2}|<1$ \\ \bottomrule
\end{tabular}%
\caption{Breakdown of the Adaptive Projection Packing method, in which a unit
  vector, $(u_x,u_y,u_z)$ is stored into two floating point numbers, FP1 and
  FP2, and one extra bit of information.}
\label{tab:adaptivepp}
\end{table}

As indicated in Table~\ref{tab:adaptivepp}, the issue of needing one extra bit
of storage space to hold the sign of the component which was projected away, has
naturally been inherited from the Static Projection method.  Although finding an
unused bit could certainly present a challenge for some applications of unit
vector packing, it is not actually an issue for the \texttt{MCPL} format where
an extra bit of information can be encoded into the sign of the floating point
number otherwise used to store the kinetic energy of the particle
(cf.\ Table~\ref{tab:mcplpart}). This bit is otherwise unused, since kinetic
energy is a non-negative quantity by definition, and is available for usage even
if particle states are specified with a kinetic energy of exactly zero. The
latter follows from the \texttt{IEEE} floating point standard~\cite{IEEE754},
which ensures availability of the sign bit even for zero (i.e.\ it provides a
\emph{signed zero} floating point implementation with distinct bit patterns for
$-0$ and $+0$). Although it is thus not needed for \texttt{MCPL}, it should be
noted for completeness that it is straight-forward to implement a variant of the
Adaptive Projection method which simply allocates the ``extra'' bit of
storage \emph{internally} in one of the two resulting floating point
numbers. This can be achieved by encoding the information into the least
significant bit of the significand of either FP1 or FP2, thus sacrificing a
small -- but hopefully inconsequential -- amount of precision in the process.

In order to compare the performance of the considered packing algorithms
quantitatively, the \emph{packing precision} when a unit vector $u$ is
transformed by the packing and unpacking into $u_p$ is defined in the
following. First, the packing precision of a single component, $i\in\{x,y,z\}$,
is defined as ${\delta}_i{\equiv}\min(1,|u_i^p/u_i|-1)$, except when $u_i=0$ in
which case it is defined to be $0$ if $u_i^p=u_i$, and $1$ otherwise. The
packing precision for the entire vector is then defined as being the worst
precision of any component, $\delta\equiv\max_i\delta_i$. The resulting values
will all lie in the unit interval and if for instance $\delta=10^{-16}$, the
packing algorithm in question can be said to have preserved the vector with at
least 16 significant digits in all components. The reason for not picking a
somewhat simpler figure of merit, such as the angular separation between $u$ and
$u^p$, is that it would be insensitive to components which are very small in
magnitude.

To be able to provide meaningful results for all considered packing methods, the
analysis presented in the following was carried out using software~\cite{mpmath}
capable of arbitrary precision floating-point arithmetic. Additionally it should
be noted that, as the focus of the present analysis is on precision and storage
consumption for a format like \texttt{MCPL}, the algorithms packing numbers into
single-precision (32 bit) floating point storage are actually implemented using
double-precision (64 bit) floating point code for all intermediate
calculations. For other particular use-cases, such as graphics rendering with
surface normals on a particular GPU, one could of course imagine also
implementing the packing code itself using single-precision everywhere.

\begin{figure}
  \centering
  \includegraphics[width=1.0\textwidth]{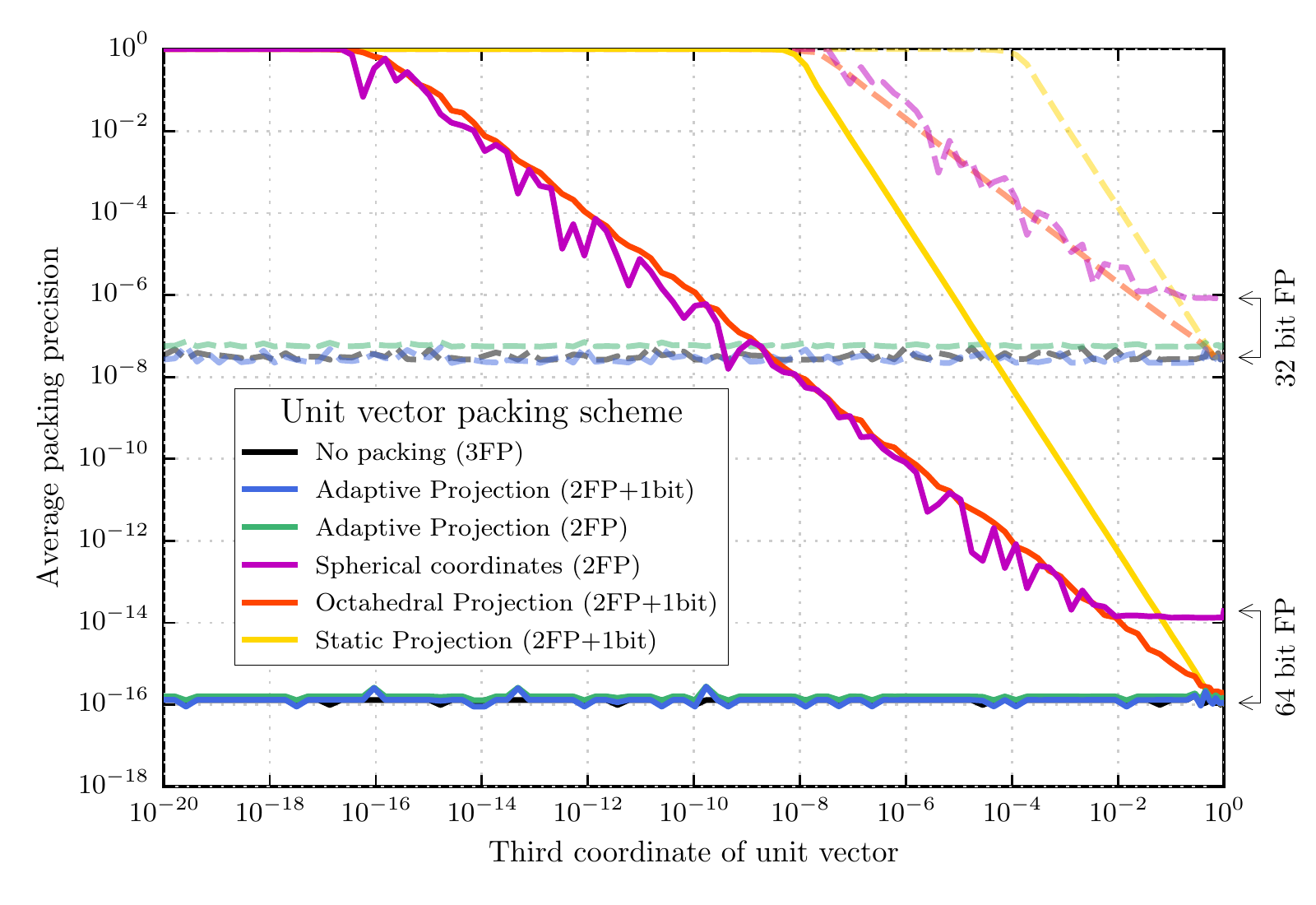}
  \caption{Average packing precision of the considered unit vector packing
    methods, as discussed in the text, for both double-precision (64 bit, full
    curves) and single-precision (32 bit, dashed opaque curves). The Adaptive
    Projection (blue, green) and No packing (black) curves are almost
    coinciding.}
  \label{fig:unitvectpacking}
\end{figure}

Motivated by the fact that the considered packing methods all have particular
issues or strategies when $u_z$ is small in magnitude,
Figure~\ref{fig:unitvectpacking} illustrates their performance when
applied to vectors with specific (possibly tiny) values of $u_z$. For each given
value of $u_z$, a test set of $10^4$ unit vectors is formed by sampling an
azimuthal angle $\phi$ uniformly in $[0,2\pi)$, and letting $(u_x,u_y)=\sqrt{1-u_z^2}\cdot(\cos\phi,\sin\phi)$.
Each packing method is then applied to the test set, and the average value of
the resulting packing precisions is plotted.  The plot clearly confirms that
while the Octahedral Projection method outperforms the Spherical coordinate and
Static Projection methods, they all degrade in performance as the magnitude of
$u_z$ tends to zero. This is qualitatively different from the Adaptive
Projection method, which shows constant performance over the entire range, at a
level which is practically indistinguishable from the case of not using any
packing -- i.e.\ storing the Cartesian coordinates directly into 3 dedicated
floating point numbers. Note that for clarity the plot is only shown here for
positive $u_z$ larger than $10^{-20}$, but it was verified in a full analysis
that, as expected, the flat level of the Adaptive Projection curves is
independent of the sign of $u_z$, and continues unchanged over the entire
dynamic range of normalised floating point numbers, down to about $10^{-38}$ and
$10^{-308}$ respectively for single- and double-precision.

As a natural figure of merit, Table~\ref{tab:uvpack_meanprecision} shows the
average packing precision of the considered methods when applied to a sample of
$10^8$ unit vectors sampled at random from an isotropic distribution. Although
all methods provide some level of precision, it is clear that only Adaptive
Projection Packing provides performance comparable to not using packing at
all. In fact, for single-precision storage, Adaptive Projection Packing in the
\emph{2FP+1bit} variant relevant for \texttt{MCPL}, even seems to be outperforming
the case of not using any packing. This somewhat counter-intuitive result is
understood to arise from the fact that the unpacking code is implemented in full
double-precision, allowing the implicit usage of the condition
$u_x^2+u_y^2+u_z^2=1$ for recovery of a component magnitude to inject a small
amount of added precision into the final result. For double-precision storage,
the precision of Adaptive Projection Packing is also seen to be practically
indistinguishable from the no packing scenario.

\begin{table}[tbp]
\centering
\begin{tabular}{@{}lllll@{}}
\toprule
\multicolumn{5}{c}{\textbf{Average packing precision in isotropic sample}}                  \\ \midrule
                               && \textit{32 bit FP} && \textit{64 bit FP}\\ \cmidrule(){3-3} \cmidrule(){5-5}
Static projection (2FP+1bit)   && $2.30\cdot10^{-4}$  && $1.76\cdot10^{-8}$            \\
Spherical coordinates (2FP)    && $1.21\cdot10^{-6}$  && $3.10\cdot10^{-15}$           \\
Octahedral packing (2FP+1bit)  && $4.11\cdot10^{-7}$  && $2.09\cdot10^{-15}$           \\
Adaptive projection (2FP)      && $5.83\cdot10^{-8}$  && $1.58\cdot10^{-16}$           \\
Adaptive projection (2FP+1bit) && $2.95\cdot10^{-8}$  && $1.20\cdot10^{-16}$           \\
No packing (3FP)               && $3.20\cdot10^{-8}$  && $1.19\cdot10^{-16}$           \\ \bottomrule
\end{tabular}%
\caption{The average observed packing precision of each considered method when
  tested on $10^8$ unit vectors sampled at random from an isotropic
  distribution, using either single- (32 bit) or double-precision (64 bit)
  floating point numbers for storage of the packed representations.}
\label{tab:uvpack_meanprecision}
\end{table}




Potentially just as crucial for users of \texttt{MCPL} as the average
performance, Table~\ref{tab:uvpack_worstprecision} shows the \emph{worst}
packing precision of any of the sampled vectors. Here, the pre-existing packing
methods, listed in the first three rows, all exhibit values far from their average
precision in Table~\ref{tab:uvpack_meanprecision}. This indicates the existence
of vectors in the sample for which the performance of the algorithms break down
-- a feature which was anticipated from the preceding discussion of flaws in
those algorithms. On the other hand, the worst encountered packing precision of the
considered Adaptive Projection Packing variants are all reasonably close to the
corresponding average values listed in Table~\ref{tab:uvpack_meanprecision}. This
once again indicates the robustness of those packing algorithms, as they indeed
do not seem to suffer from breakdown on certain domains.

\begin{table}[tbp]
\centering
\begin{tabular}{@{}lllll@{}}
\toprule
\multicolumn{5}{c}{\textbf{Worst packing precision in isotropic sample}}  \\ \midrule
                               && \textit{32 bit FP} && \textit{64 bit FP}\\ \cmidrule(){3-3} \cmidrule(){5-5}
Static projection (2FP+1bit)   && \hspace*{1.9em}$1$        && \hspace*{1.9em}$1$           \\
Spherical coordinates (2FP)    && \hspace*{1.9em}$1$        && $1.12\cdot10^{-8}$           \\
Octahedral packing (2FP+1bit)  && \hspace*{1.9em}$1$        && $9.51\cdot10^{-9}$           \\
Adaptive projection (2FP)      && $2.03\cdot10^{-7}$            && $7.50\cdot10^{-16}$          \\
Adaptive projection (2FP+1bit) && $1.01\cdot10^{-7}$            && $5.77\cdot10^{-16}$          \\
No packing (3FP)               && $5.96\cdot10^{-8}$            && $2.22\cdot10^{-16}$          \\ \bottomrule
\end{tabular}%
\caption{The worst observed packing precision of each considered method when
  tested on $10^8$ unit vectors sampled at random from an isotropic
  distribution, using either single- (32 bit) or double-precision (64 bit)
  floating point numbers for storage of the packed representations.}
\label{tab:uvpack_worstprecision}
\end{table}




In summary, the presented Adaptive Projection Packing method and in particular
the \emph{2FP+1bit} variant adopted for \texttt{MCPL}, provides a packing
precision which is for all practical purposes comparable to that given by three
floating point numbers. The difference in performance is so small that it is
hard to imagine use-cases which would be satisfied with the latter but not by
the former. For that reason, it was decided to not add a no-packing option to
\texttt{MCPL} at this point.

\vfill\clearpage

\section{\texttt{C}-API for \texttt{MCPL} files}\label{appendix:reference_c_api}

For reference, this appendix documents the available functions, data structures
and constants in the API exposed in \texttt{mcpl.h} as of \texttt{MCPL} version
1.1.0. The file and its contents can be directly included and used from code
compiled with any official standard of either \texttt{C} or \texttt{C++}.

\subsection{Data structures and constants}\label{appendix:c_api_dataandconsts}

\begin{docfct}
#define MCPL_VERSION_MAJOR 1
#define MCPL_VERSION_MINOR 1
#define MCPL_VERSION_PATCH 0
#define MCPL_VERSION   10100
#define MCPL_VERSION_STR "1.1.0"
\end{docfct}%
\fctdesc{ Pre-processor macros providing information about the version of MCPL.
  They should hopefully be self-explanatory except perhaps \fctname{MCPL_VERSION} which
  encode the version into a single integer,
  10000*MAJOR+100*MINOR+PATCH, allowing easy comparison of versions numbers.
}

\begin{docfct}
#define MCPL_FORMATVERSION 3
\end{docfct}%
\fctdesc{ Pre-processor macro providing the file format version of files written
  by the installation of \texttt{MCPL}. Note that this file format version
  (currently 3) is not the same as the version number of the \texttt{MCPL}
  distribution (currently 1.1.0), and the latter is expected to be updated more
  frequently. The current third version of the file format is described in
  detail in \ref{appendix:mcpldetailedlayout}, and the file format of a given
  \texttt{MCPL} file can be queried through the function
  \fctname{mcpl_hdr_version} described in \ref{appendix:c_api_filereading}.  }

\begin{docfct}
  typedef struct {
    double ekin;            /* kinetic energy [MeV]             */
    double polarisation[3]; /* polarisation vector              */
    double position[3];     /* position [cm]                    */
    double direction[3];    /* momentum direction (unit vector) */
    double time;            /* time-stamp [millisecond]         */
    double weight;          /* weight or intensity              */
    int32_t pdgcode;
    uint32_t userflags;
  } mcpl_particle_t;
\end{docfct}%
\fctdesc{ Data structure representing a particle. Pointers to
  \fctname{mcpl_particle_t} instances are respectively returned from and passed
  to \fctname{mcpl_read} (cf.~\ref{appendix:c_api_filereading}) and
  \fctname{mcpl_add_particle} (cf.~\ref{appendix:c_api_filecreation}) when
  extracting or adding particles. Refer to the description of those functions
  for further details.
}

\begin{docfct}
  typedef struct { void * internal; } mcpl_file_t
  typedef struct { void * internal; } mcpl_outfile_t
\end{docfct}%
\fctdesc{ Data structures representing file handles to \texttt{MCPL}
  files. Typically file handles are created and returned by a call to either
  \fctname{mcpl_open_file} or \fctname{mcpl_create_outfile} as appropriate, and
  functions performing subsequent operations on the file all take the file
  handle as one of the arguments. Note that the size of these structures is
  identical to that of a pointer, and they are therefore appropriate to pass or
  return by value from functions with no special overhead. The \fctname{void *
    internal} pointer is, as the name implies, for purely internal usage in
  \texttt{mcpl.c}. Client code should never access or modify this pointer in any
  way, except perhaps to set it to NULL in order to mark a file handle as
  uninitialised or invalid. Refer to \ref{appendix:c_api_filecreation} and
  \ref{appendix:c_api_filereading} for further details of how these file handles
  can be used.}

\subsection{Functions for file creation}\label{appendix:c_api_filecreation}

Creation of new \texttt{MCPL} files always begins with a call to
\fctname{mcpl_create_outfile}, returning a file handle of the type
\fctname{mcpl_outfile_t} (cf.~\ref{appendix:c_api_dataandconsts}). This file
handle is then passed in to various functions in order to first set flags and
meta-data and then add particles with \fctname{mcpl_add_particle}. Finally, the
file must be properly closed with a call to either \fctname{mcpl_close_outfile}
or \fctname{mcpl_closeandgzip_outfile}.

\begin{docfct}
mcpl_outfile_t mcpl_create_outfile(const char * filename)
\end{docfct}%
\fctdesc{Function used to start creation of a new \texttt{MCPL} file.  It
  attempts to open the indicated file for writing (overriding any existing file)
  and returns a handle to the caller. The latter must subsequently be passed in
  as a parameter to other functions below, in order to configure the file, add
  particles to it, and ultimately close it.  Note that if the provided file name
  does not end with the extension \texttt{``.mcpl''}, it will be automatically
  appended to it (see also \fctname{mcpl_outfile_filename} below).}

\begin{docfct}
const char * mcpl_outfile_filename(mcpl_outfile_t)
\end{docfct}%
\fctdesc{Filename being written to. If the filename passed to
  \fctname{mcpl_create_outfile} ended with ``.mcpl'', it will be
  identical to what is returned. Otherwise, a postfix of ``.mcpl'' will have been
  appended.}

\gdef\stdsentencecreatehdr{ This function must be called before any calls to  \fctname{mcpl_add_particle}. }

\begin{docfct}
void mcpl_hdr_set_srcname(mcpl_outfile_t, const char *)
\end{docfct}%
\fctdesc{Optionally set name of the generating application which will be stored in
  the file header. If not called, a string with content ``unknown'' will be
  stored instead. \stdsentencecreatehdr }

\begin{docfct}
void mcpl_hdr_add_comment(mcpl_outfile_t,const char *)\end{docfct}%
\fctdesc{Add one or more human-readable comments to the file.  \stdsentencecreatehdr}

\begin{docfct}
void mcpl_hdr_add_data(mcpl_outfile_t, const char * key,
                       uint32_t ldata, const char * data)
\end{docfct}%
\fctdesc{Add a binary data ``blob'' and associate it with a given key, which
  must be unique to the file. \stdsentencecreatehdr}

\begin{docfct}
void mcpl_enable_userflags(mcpl_outfile_t)
\end{docfct}%
\fctdesc{Enable per-particle user-flags to be written in the file. If not called,
  any non-zero value in the \fctname{userflags} field of added particles
  will be ignored by \fctname{mcpl_add_particle}. \stdsentencecreatehdr}

\begin{docfct}
void mcpl_enable_polarisation(mcpl_outfile_t)
\end{docfct}%
\fctdesc{Enable per-particle polarisation vectors to be written in the file. If not called,
  any non-zero values in the \fctname{polarisation} field of added particles
  will be ignored by \fctname{mcpl_add_particle}. \stdsentencecreatehdr}

\begin{docfct}
void mcpl_enable_doubleprec(mcpl_outfile_t)
\end{docfct}%
\fctdesc{Enables double-precision (64 bit) storage of floating point numbers in
  particle data. Default is otherwise single-precision (32 bit). \stdsentencecreatehdr}

\begin{docfct}
void mcpl_enable_universal_pdgcode(mcpl_outfile_t, int32_t pdgcode)
\end{docfct}%
\fctdesc{Prevent per-particle PDG codes from being written in the file, letting
  instead all particles have the same common code. This means that
  values in the \fctname{pdgcode} field of added particles
  will be ignored by \fctname{mcpl_add_particle}. \stdsentencecreatehdr}

\begin{docfct}
void mcpl_enable_universal_weight(mcpl_outfile_t, double uw)
\end{docfct}%
\fctdesc{Prevent per-particle weights from being written in the file, letting
  instead all particles have the same common weight. This means that
  values in the \fctname{weight} field of added particles
  will be ignored by \fctname{mcpl_add_particle}.
  Typical usage is to save
  one floating point of per-particle storage, when simulation output is known to
  not be weighted, by setting a universal weight of \fctname{1.0}.
  \stdsentencecreatehdr}

\begin{docfct}
void mcpl_add_particle(mcpl_outfile_t,const mcpl_particle_t*)
\end{docfct}%
\fctdesc{Add the particle state represented by the values on the provided
  \fctname{mcpl_particle_t} instance, to the file. Note that some fields on the
  \fctname{mcpl_particle_t} instance might be ignored, depending on the settings
  of the file (cf.\ the preceding descriptions of functions altering the file
  settings). Note that after this function has been called, the header is
  flushed to disk and file settings can consequently no longer be modified --
  attempting to do so will result in an error. Also note that the
  \fctname{mcpl_get_empty_particle} function described below provides a
  convenient manner to obtain a \fctname{mcpl_particle_t} instance to modify and pass to
  this function, without having to worry about memory management.}

\begin{docfct}
void mcpl_close_outfile(mcpl_outfile_t)
\end{docfct}%
\fctdesc{Close file and flush all pending data to disk, thus representing the
  last but non-optional step in the creation of a new \texttt{MCPL} file. It is
  undefined behaviour to attempt to use a file handle for anything after passing it to this function.
}

\begin{docfct}
int mcpl_closeandgzip_outfile(mcpl_outfile_t)
\end{docfct}%
\fctdesc{A convenience function which first calls \fctname{mcpl_close_outfile}
  and then \fctname{mcpl_gzip_file} (cf.~\ref{appendix:c_api_otherfcts}),
  forwarding the return value of the latter. This means that, on platforms where
  this is possible, the function will return 1 to indicate that
  the newly created \texttt{MCPL} file will have been compressed with
  \texttt{GZIP} and the final output file will have had ``.gz'' appended to its
  name. If the function returns 0, the \texttt{MCPL} file will still have been
  created, but it will not have been compressed with \texttt{GZIP}.}

\begin{docfct}
mcpl_particle_t* mcpl_get_empty_particle(mcpl_outfile_t)
\end{docfct}%
\fctdesc{Convenience
  function which returns a pointer to a nulled-out particle struct which can be
  used to edit and pass to \fctname{mcpl_add_particle}. It can be reused and will be
  automatically deallocated when the file is closed.}

\subsection{Functions for reading file data}\label{appendix:c_api_filereading}

Access to data in existing \texttt{MCPL} files always begins with a call to
\fctname{mcpl_open_file}, returning a file handle of the type
\fctname{mcpl_file_t} (cf.~\ref{appendix:c_api_dataandconsts}). This file handle
is then passed in to various functions in order to either access file settings
and meta-data, or particle data with the \fctname{mcpl_read} function. The
latter function reads the particle at the current position and skips forward to
the next one, but functions of course also exist which facilitate skipping and
seeking to any position in the file. Finally, to ensure release of all
resources, the file handle should be released with a call to
\fctname{mcpl_close_file}.

\begin{docfct}
mcpl_file_t mcpl_open_file(const char * filename);
\end{docfct}%
\fctdesc{ Open indicated file and load header information into memory, skip to
  the first (if any) available particle.  When \texttt{ZLIB} support is enabled
  (cf.\ section~\ref{sec:buildanddeploy}), this function can read compressed
  \texttt{MCPL} files with the \texttt{.mcpl.gz} extension directly.}

\begin{docfct}
  unsigned mcpl_hdr_version(mcpl_file_t)
\end{docfct}%
\fctdesc{
  File format version of the \texttt{MCPL} file. See the description of
  the \fctname{MCPL_FORMATVERSION} macro in \ref{appendix:c_api_dataandconsts}
  for more details.}

\begin{docfct}
  uint64_t mcpl_hdr_nparticles(mcpl_file_t)
\end{docfct}%
\fctdesc{
  Number of particles stored in the file.
}

\begin{docfct}
  const char* mcpl_hdr_srcname(mcpl_file_t)
\end{docfct}%
\fctdesc{  Name of the generating application.}

\begin{docfct}
  unsigned mcpl_hdr_ncomments(mcpl_file_t)
\end{docfct}%
\fctdesc{
  Number of comments stored in the file.
}

\begin{docfct}
  const char * mcpl_hdr_comment(mcpl_file_t, unsigned i)
\end{docfct}%
\fctdesc{ Access \fctname{i}'th comment stored in the file. If \fctname{i} is not a number smaller
  than the number of comments in the file (cf.~\fctname{mcpl_hdr_ncomments}), an error results.  }

\begin{docfct}
  int mcpl_hdr_nblobs(mcpl_file_t);
\end{docfct}%
\fctdesc{
  Number of binary data ``blobs'' stored in the file.
}

\begin{docfct}
const char** mcpl_hdr_blobkeys(mcpl_file_t)
\end{docfct}%
\fctdesc{ Returns a list of the binary data keys available in the file. The
  function returns NULL if there are no binary data ``blobs'' in the file.  }

\begin{docfct}
  int mcpl_hdr_blob(mcpl_file_t, const char* key,
                    uint32_t* ldata, const char ** data)
\end{docfct}%
\fctdesc{
  Access the binary data array associated with a given key. The function returns
  0 if the key does not exist in the file (cf.~\fctname{mcpl_hdr_blobkeys}). Otherwise it returns 1
  and \fctname{ldata} and \fctname{data} will have been modified to
  contain respectively the length of the data and the address of the data.
}

\begin{docfct}
  int mcpl_hdr_has_userflags(mcpl_file_t);
\end{docfct}%
\fctdesc{Returns 1 if per-particle user-flags are stored in the file. If not, 0
  is returned and all particles read from the file will have a \fctname{userflags} field
  value of \fctname{0x0}.}

\begin{docfct}
  int mcpl_hdr_has_polarisation(mcpl_file_t);
\end{docfct}%
\fctdesc{Returns 1 if per-particle polarisation vectors are stored in the
  file. If not, 0 is returned and all particles read from the file will have a
  null vector in the \fctname{polarisation} field.}

\begin{docfct}
  int mcpl_hdr_has_doubleprec(mcpl_file_t);
\end{docfct}%
\fctdesc{Returns 1 if floating point numbers in the particle data in the file is
  stored using double-precision (64 bit) as opposed to single-precision (32 bit)
  numbers.  }

\begin{docfct}
  uint64_t mcpl_hdr_header_size(mcpl_file_t);
\end{docfct}%
\fctdesc{
  Returns the number of bytes consumed by the header on disk (uncompressed).
}

\begin{docfct}
  int mcpl_hdr_particle_size(mcpl_file_t);
\end{docfct}%
\fctdesc{
  Returns the number of bytes consumed by each particle on disk (uncompressed).
}

\begin{docfct}
  int32_t mcpl_hdr_universal_pdgcode(mcpl_file_t);
\end{docfct}%
\fctdesc{Returns zero if per-particle PDG codes are stored in the file. If not,
  the returned value is the common value which all particles read from the
  file will have in the \fctname{pdgcode} field.}

\begin{docfct}
  double mcpl_hdr_universal_weight(mcpl_file_t);
\end{docfct}%
\fctdesc{Returns zero if per-particle weights are stored in the file. If not,
  the returned value is the common value which all particles read from the
  file will have in the \fctname{weight} field.}

\begin{docfct}
  int mcpl_hdr_little_endian(mcpl_file_t);
\end{docfct}%
\fctdesc{Returns 1 if the numbers in the file are stored in little-endian form,
  and 0 if they are stored in big-endian form.}

\begin{docfct}
  const mcpl_particle_t* mcpl_read(mcpl_file_t);
\end{docfct}%
\fctdesc{Attempts to read the particle at the current position in file and skips
  forward to the next particle. Return value will be NULL in case there was no
  particle at the current location (normally due to end-of-file), otherwise it
  will be a pointer to an \fctname{mcpl_particle_t} instance representing the particle
  just read. Note that the returned pointer is invalidated if the file is closed
  or \fctname{mcpl_read} is called again.}

\begin{docfct}
  uint64_t mcpl_currentposition(mcpl_file_t);
\end{docfct}%
\fctdesc{
  Returns current position in the file, which is a number less than or equal to
  the number of particles in the file, $N$
  (cf.~\fctname{mcpl_hdr_nparticles}). If $N$ is returned, this indicates an
  end-of-file condition where no more particles can be read with
  \fctname{mcpl_read}.
}

\begin{docfct}
  int mcpl_skipforward(mcpl_file_t,uint64_t n);
\end{docfct}%
\fctdesc{ Advance position in file \fctname{n} steps. Returns 0 if this causes
  the end-of-file to be reached and there is no particle to be read at the new
  position. Otherwise returns 1. }

\begin{docfct}
  int mcpl_rewind(mcpl_file_t);
\end{docfct}%
\fctdesc{ Rewinds position in file to 0. Returns 0 if this causes the
  end-of-file to be reached and there is no particle to be read at the new
  position (under normal conditions, this should only happen for empty files
  with no particles). Otherwise returns 1. }

\begin{docfct}
  int mcpl_seek(mcpl_file_t,uint64_t ipos);
\end{docfct}%
\fctdesc{ Seek directly to specified position in file. Returns 0 if this causes
  the end-of-file to be reached and there is no particle to be read at the new
  position. Otherwise returns 1.}

\begin{docfct}
  void mcpl_close_file(mcpl_file_t);
\end{docfct}%
\fctdesc{
  Deallocate memory and release file-handle.
  It is undefined behaviour to attempt to use a file handle for anything after passing it to this function.
}

\subsection{Other functions}\label{appendix:c_api_otherfcts}

\begin{docfct}
void mcpl_dump(const char * file, int parts, uint64_t nskip, uint64_t nlimit)
\end{docfct}%
\fctdesc{Prints information about the specified \texttt{MCPL} file to standard
  output. This is similar to what is printed by the \texttt{mcpltool} at the
  command line. The parameter \fctname{parts} can be used to control whether or
  not to print information from just the file header (\fctname{parts=1}), just
  the particle state data (\fctname{parts=2}) or both (\fctname{parts=0}). If
  particle state data is listed, \fctname{nskip} and \fctname{nlimit} can be
  used to control which of the contained particles to list: \fctname{nlimit} is
  an upper limit on the number of particles printed (\fctname{nlimit=0} means no
  limit), and \fctname{nskip} can be used to skip that many positions into the
  file before starting the printouts.}

\begin{docfct}
mcpl_outfile_t mcpl_merge_files(const char* file_output,
                                unsigned nfiles, const char ** files)
\end{docfct}%
\fctdesc{Merge contents of a list of files by concatenating all particle
  contents into a new output file. This results in an error
  unless all meta-data and settings in the files are identical. For safety, this
  fails if \fctname{file_output} already exists. The function returns a handle
  to the output file which has had all particles added to it, but has not yet been
  closed. Note that if any file is specified more than once in the input list, a
  warning will be printed to standard output.}

\begin{docfct}
int mcpl_can_merge(const char * file1, const char* file2)
\end{docfct}%
\fctdesc{
  Test if files could be merged by \fctname{mcpl_merge_files}. This returns 1 if
  all meta-data and settings in the files are identical, otherwise 0.
}

\begin{docfct}
void mcpl_merge_inplace(const char * file1, const char* file2);
\end{docfct}%
\fctdesc{ Similar to \fctname{mcpl_merge_files}, but merges two files by
  appending all particles in \fctname{file2} to the list in \fctname{file1}
  (thus \fctname{file1} grows while \fctname{file2} stays untouched).  Note that
  this requires similar version of the file format of the two files, in addition
  to the other checks in \fctname{mcpl_can_merge}. Careful usage of this
  function can be more efficient than \fctname{mcpl_merge_files}, but it is
  potentially also less safe as \fctname{file1} is left modified. Note that if
  \fctname{file1} and \fctname{file2} are the same file, a warning will be
  printed to standard output.}

\begin{docfct}
void mcpl_repair(const char * file1)
\end{docfct}%
\fctdesc{ Attempts to repair a broken file which was never properly closed. This
  is intended for recovery of contents in files produced in long jobs which were
  interrupted for some reason, and thus never had the number of particles field
  in the file header updated correctly. It works by using the file size to
  calculate the number of complete particle entries in the file, and then
  updating the header.}

\begin{docfct}
int mcpl_tool(int argc, char** argv)
\end{docfct}%
\fctdesc{ This function implements the command line \texttt{mcpltool} command,
  and should be wrapped inside a standard \texttt{main} function of a \texttt{C}
  application, which should then be compiled into an executable named
  \texttt{mcpltool} or similar.  }

\begin{docfct}
  int mcpl_gzip_file(const char * filename);
\end{docfct}%
\fctdesc{ Attempts to compress file with \texttt{GZIP}, appending ``.gz'' to its
  name in the process. On platforms where this is possible, the function will
  return 1 to indicate success. If the function instead returns 0, the
  file will have been left untouched.}

\begin{docfct}
void mcpl_transfer_metadata(mcpl_file_t source, mcpl_outfile_t target);
\end{docfct}%
\fctdesc{
   Convenience function which transfers all settings, blobs and comments to
   outfile. Intended to make it easy to filter files via custom C code.
}

\begin{docfct}
  void mcpl_set_error_handler(void (*handler)(const char *));
\end{docfct}%
\fctdesc{ Override the default \texttt{MCPL} error handler, which is a function
  that will get called with a string describing the error and which should never
  return to the calling code. If no handler is set, errors will get printed to
  standard output and the process terminated.}

\vfill\clearpage

\section{Usage instructions for \texttt{mcpltool}}\label{appendix:reference_mcpltool_usage}

Usage instructions of the \texttt{mcpltool} command are available from the
command line by invoking it with the \texttt{-{}-help} flag. For reference the
output is repeated here in Listing~\ref{lst:mcpltoolusage}.

\vspace*{2ex}

\begin{lstlisting}[language={},
  basicstyle={\linespread{0.9}\ttfamily\ssmall},
  label={lst:mcpltoolusage},
  caption={Usage instructions for the \texttt{mcpltool} command (output of ``\texttt{mcpltool~-{}-help}'').}
]
Tool for inspecting or modifying Monte Carlo Particle List (.mcpl) files.

The default behaviour is to display the contents of the FILE in human readable
format (see Dump Options below for how to modify what is displayed).

This installation supports direct reading of gzipped files (.mcpl.gz).

Usage:
  ess_mcpl_tool [dump-options] FILE
  ess_mcpl_tool --merge [merge-options] FILE1 FILE2
  ess_mcpl_tool --extract [extract-options] FILE1 FILE2
  ess_mcpl_tool --repair FILE
  ess_mcpl_tool --version
  ess_mcpl_tool --help

Dump options:
  By default include the info in the FILE header plus the first ten contained
  particles. Modify with the following options:
  -j, --justhead  : Dump just header info and no particle info.
  -n, --nohead    : Dump just particle info and no header info.
  -lN             : Dump up to N particles from the file (default 10). You
                    can specify -l0 to disable this limit.
  -sN             : Skip past the first N particles in the file (default 0).
  -bKEY           : Dump binary blob stored under KEY to standard output.

Merge options:
  -m, --merge FILEOUT FILE1 FILE2 ... FILEN
                    Creates new FILEOUT with combined particle contents from
                    specified list of N existing and compatible files.
  -m, --merge --inplace FILE1 FILE2 ... FILEN
                    Appends the particle contents in FILE2 ... FILEN into
                    FILE1. Note that this action modifies FILE1!

Extract options:
  -e, --extract FILE1 FILE2
                    Extracts particles from FILE1 into a new FILE2.
  -lN, -sN          Select range of particles in FILE1 (as above).
  -pPDGCODE         select particles of type given by PDGCODE.

Other options:
  -r, --repair FILE
                    Attempt to repair FILE which was not properly closed, by up-
                    dating the file header with the correct number of particles.
  -v, --version   : Display version of MCPL installation.
  -h, --help      : Display this usage information (ignores all other options).
\end{lstlisting}

\end{document}